\documentclass[aps,prl,twocolumn,10pt,svgnames]{revtex4-2}
\usepackage{amsmath,amssymb,amsthm}
\usepackage{amsmath,amssymb,mathtools}
\usepackage{hyperref}
\usepackage[capitalize,nameinlink,noabbrev]{cleveref}
\usepackage{graphicx,braket,physics,bm,hyperref}
\usepackage{listings}
\usepackage{tikz}
\usepackage{braket}
\usepackage{quantikz} 
\usepackage[toc,page]{appendix}
\usepackage{enumerate}
\usepackage{physics}
\usepackage{tikz}
\usepackage{siunitx}
\usetikzlibrary{arrows.meta, positioning}
\usepackage{quantikz}
\usepackage{hyperref}
\usepackage{amsthm} 
\theoremstyle{plain}
\newtheorem{lemma}{Lemma}[section]

\theoremstyle{definition}    
\newtheorem{definition}{Definition}[section]
\theoremstyle{remark}        
 
\theoremstyle{construction}        
 
\theoremstyle{proposition}        
\newtheorem*{proposition}{Proposition} 
\crefname{section}{Sec.}{Secs.}
\Crefname{section}{Section}{Sections}
\crefname{subsection}{Sec.}{Secs.}
\Crefname{subsection}{Section}{Sections}
\crefname{theorem}{Thm.}{Thms.}
\Crefname{theorem}{Theorem}{Theorems}
\crefname{proposition}{Prop.}{Props.}
\Crefname{proposition}{Proposition}{Propositions}
\crefname{appendix}{App.}{Apps.}
\crefname{equation}{Eq.}{Eqs.}
\usepackage{amsmath}
\usepackage{subcaption}
\usepackage{hyperref}     
\usepackage{physics}

\usepackage{enumitem}
\usepackage{amsfonts}
\newcommand{\suchthat}{\;\ifnum\currentgrouptype=16 \middle\fi|\;}

\providecommand{\Tr}{\operatorname{Tr}}

\providecommand{\E}{\mathbb{E}}
\providecommand{\Pr}{\mathbb{P}}
\providecommand{\RR}{\mathbb{R}}

\providecommand{\Id}{\mathrm{I}}
\providecommand{\1}{\mathbf{1}}

\begin{document}

\title{Random-projector quantum diagnostics of Ramsey numbers and a prime-factor heuristic for $R(5,5)=45$}

\author{Fabrizio Tamburini} 
\email{fabrizio.tamburini@gmail.com (Erd\H{o}s number = 5)}
\affiliation{Rotonium -- Quantum Computing, Le Village by CA, Piazza G. Zanellato, 23, 35131 Padova PD, Italy.}
\date{\today}

\begin{abstract}
We introduce a statistical framework for estimating Ramsey numbers by embedding two-color Ramsey instances into a $\mathbb{Z}_2 \times \mathbb{Z}_2$-graded Majorana algebra. This approach replaces brute-force enumeration with two randomized spectral diagnostics applied to operators of a given dimension $d$ associated with Ramsey numbers: a linear projector $P_{\mathrm{lin}}$ and an exponential map $P_{\exp}(\alpha)$, suitable for both classical and quantum computation. In the diagonal case, both diagnostics identify $R(5,5)$ at $n=45$. The quantum realizations act on a reduced module and therefore require only five data qubits plus a few ancillas via block-encoding/qubitization for $R(5,5)=45$, in stark contrast to the $\binom{n}{2} \approx 10^3$ logical qubits demanded by direct edge encodings. We also provide few-qubit estimates for $R(6,6)$ and $R(7,7)$, and propose a simple ``prime-sequence'' consistency heuristic that connects $R(5,5)=45$ to constrained diagonal growth. Our method echoes Erd\H{o}s's probabilistic paradigm, emphasizing randomized arguments rather than explicit colorings, and parallels the classical coin-flip approach to Ramsey bounds. Finally, we discuss potential applications of this framework to machine learning with a limited number of qubits.
\end{abstract}

\maketitle

\section{Introduction}
Ramsey numbers are at the intersection of graph theory, combinatorics, neural networking, computation and probability. Ramsey theory asks for the smallest number $R(m,n)$ such that every edge coloring of the complete graph $K_{R(m,n)}$ with two colors contains a red graph $K_m$ with $m$ vertices or a blue $K_n$ with $n$ vertices \cite{Ramsey1930}. In graph theory, a Ramsey number refers to the minimum number of vertices in a complete graph where, regardless of how the edges are colored with two colors, a monochromatic subgraph of a specified size (or sizes) is guaranteed to exist. Ramsey’s theorem guarantees the number exists. We know all exact values up to $R(4,5)=25$; the next diagonal case $43<R(5,5)\leq 46$ has not been directly calculated yet \cite{exoo,angel,angel1}.
Gap sizes grow explosively between $R(5,5)$ and $R(6,6)$ the interval already spans $60$ numbers, and for $R(7,7)$ we only know $189 \leq R(7,7) \leq 4749$. It is the search of sequences and patterns in large structures like searching for constellations in the sky \cite{sciamerican}.

Ramsey numbers control worst‑case resource overheads in error‑correcting codes and result intractable with current quantum computing resources. Designing a code to correct $t$ errors means ensuring no two codewords sit within Hamming distance, which measures the minimum number of substitutions required to change one string into the other, $d_H \leq 2t$ \cite{hamming}. This avoidance problem is equivalent to a set-coloring Ramsey instance on an alphabet of size $q$. 
The threshold at which large codes cease to exist scales like $R(t+1; q, q-1)$, so the redundancy one must pay in the worst case grows with a Ramsey number rather than a mere polynomial in $t$. This follows from the standard avoidance formulation of code design; see  \cite{Alon2004} for the probabilistic method and \cite{GrahamSpencerRamsey} for Ramsey-theoretic background.

In quantum computing they are fundamental in worst-case routing in quantum compilation, mapping an arbitrary $k$-qubit circuit onto limited hardware connectivity forces SWAP gates whenever interacting qubits aren’t adjacent. 
Ramsey’s theorem guarantees that in any coloring or pattern of two-qubit interactions among $R(k,k)$ qubits, there must exist an unavoidable “hard” configuration such as a clique or independent set requiring at least $R(k,k)$ SWAP operations or depth units to resolve. Thus, the diagonal Ramsey number sets a lower bound on the worst-case communication depth of any fully general distributed quantum compiler in worst-case interaction patterns under limited connectivity, a Ramsey-style obstruction yields a depth lower bound scaling with a diagonal Ramsey number~\cite{GrahamSpencerRamsey,Childs2019Circuit}.
To push worst-case depths down, one either needs richer native connectivity or higher-dimensional qudits of order $n$ to effectively reduce the Ramsey argument to $R([k/n],[k/n])$ adopting paraparticle methods with  $\mathbb Z_{2} \times \mathbb Z_{2}$--graded Majorana algebras \cite{tambu1}.

Computing Ramsey numbers is notoriously intensive: classical proofs use double counting or induction \cite{rads}, whereas the gluing technique builds larger graphs from smaller ones to avoid or force specific subgraphs such as cliques or independent sets \cite{goed}.

We propose a new statistical method to estimate Ramsey number based on $\mathbb Z_2 \times \mathbb Z_2$-graded Lie Algebras of Majorana infinite-component spin fields introduced in \cite{tambu1}. 
Our strategy follows the implementing of a paraparticle algebra graded by the Klein group $V_4\cong\mathbb Z_2\times\mathbb Z_2$, which requires only a limited number of qubits for each step of analysis. By implementing higher-order algebras we develop recursive gluing and pruning methods on Ramsey graphs to give an estimate to or effectively calculate Ramsey numbers.
The result is a purely algebraic that takes in account of the branching that factorizes the two colors and packages local symmetry constraints, allowing traces/character formulas to replace brute-force enumeration for many branches.

Quantum resources needed are formidable: verifying $R(4,4)=18$ requires exhaustively scanning all $2^{136}$ two--colourings of $K_{17}$, i.e.\ one logical qubit per edge $(136$ qubits$)$, already at the frontier of current hardware~\cite{wang,gaitan}.  Even with Grover’s quadratic speed-up the bottleneck is width, not depth: a 40-vertex instance occupies a $2^{780}$-dimensional space, demanding 780 qubits before amplification can begin.

For $R(5,5)$ we give an estimate of the qubit cost of a brute--force (Grover--style) search that needs an enormous number of quantum resources with a register model based on a two--colouring of $K_n$ that can be encoded by a binary string $x\in\{0,1\}^{E(n)}$ with one bit per (undirected) edge, where
$E(n)=\binom{n}{2}=n(n-1)/2$.
Adopting the convention $x_{ij}=1$ \textit{iff} edge $\{i,j\}$ is red (blue is $0$), a reversible predicate $F_n(x)$ flags a violation ($F_n(x)=1$) \textit{iff} $x$ contains a red $K_5$ or a blue $K_5$; otherwise $F_n(x)=0$.  
In Grover’s search for a good colouring one uses the oracle
\begin{equation}
\mathcal{O}_n:\ |x\rangle|b\rangle\ \mapsto\ |x\rangle\,|b\oplus \neg F_n(x)\rangle,
\end{equation}
or phase kickback with a $|-\rangle$ ancilla.

Oracle structure and ancillas can be organized in this way, like a game: for each 5--subset $S\subseteq[n]$ we must test whether all $10=\binom{5}{2}$ edge--bits inside $S$ are equal to $1$ (red $K_5$) or all equal to $0$ (blue $K_5$).
Each “all--red” test is a 10--input AND operation, which can be realized by a Toffoli tree using $9$ clean ancillas (one per internal node); the “all--blue” test is obtained by negating the same $10$ inputs, reusing the same $9$ ancillas, and then uncomputing.  
Let $a_{\mathrm{AND}}=9$ denote these workspace ancillas. We keep one flag $r_S$ for “red $K_5$ on $S$”, one flag $b_S$ for “blue $K_5$ on $S$”, and an aggregate OR--bit $v$ that accumulates violations via $v\leftarrow v\lor (r_S\lor b_S)$ as we sweep all $\binom{n}{5}$ subsets, uncomputing $r_S,b_S$ and the AND tree at the end of each subset.
Finally, we need one output/phase ancilla for Grover, $a_{\mathrm{ph}}=1$.

Thus the oracle’s \emph{simultaneous} ancilla budget can be kept to have a number of ancillas, 
$a_{\text{work}} = a_{\mathrm{AND}} + 2+1+a_{\mathrm{ph}}
= 9 + 2 + 1 + 1 = 13$, i.e., a number of $13$ clean ancillas beyond the edge register. If, instead, negative controls are disallowed and input--negations must be done explicitly for the “all--blue” check, one adds at most $10$ transient bits that are immediately uncomputed; the peak simultaneous ancillas remain $\le 13$.  One may also think to add a couple of safety ancillas for carry/OR trees; which means a budget of $+3$ that would be needed to cover such variants, but it ia a matter of optimization that goes beyond the purpose of this work.

The total qubits required starts from a conservative upper bound $Q(n)$ on the width (qubit count) of a brute--force Grover oracle for the diagonal $R(5,5)$ tests at size $n$,
\begin{equation}
Q(n)\;=\;E(n) \;+\; a_{\text{work}} \;\;\lesssim\;\; \binom{n}{2} + 16.
\end{equation}
Concretely, as reported in the following table, the number of qubit required is very big
\begin{table}[h]
\caption{Estimate of the number of qubits required to calculate $R(5,5)$ with Grover-style algorithm}
\begin{tabular}{c|c|c}
$n$ & $E(n)=\binom{n}{2}$ & Total qubits $Q(n)$ \\ 
 &  & (\text{safe upper bound}) \\ 
\hline
$44$ & $44\cdot 43/2 = 946$  & $946 + 16 \;\approx\; 962$ \\
$45$ & $45\cdot 44/2 = 990$ & $990 + 16 \;\approx\; 1006$ \\
$46$ & $46\cdot 45/2 = 1035$ & $1035 + 16 \;\approx\; 1051$
\end{tabular}
\label{}
\end{table}

The search space has size $2^{E(n)}=2^{\binom{n}{2}}$.  Grover’s amplitude amplification finds a good colouring, when one exists, n has $O\!\big(\sqrt{2^{E(n)}/G_n}\big)$ oracle uses, where $G_n$ is the number of good colourings at size $n$, which is unknown a priori. For the upper--bound instance (let us assume $n=45$), proving nonexistence of a good colouring via search still takes exponential time in the worst case and requires additional outer logic, but the width remains dominated by the edge register as above.

The truly brute--force (enumeration/Grover) approach to $R(5,5)$ at the threshold requires on the order of a thousand logical qubits just to hold a colouring ($\binom{45}{2}=990$), plus $\lesssim 13+3=16$ clean ancillas for the reversible violation check and phase kickback.  

This stands in sharp contrast to our estimate random--projector spectral diagnostics presented in the next section, whose quantum implementation needs only a few data qubits (plus a handful of ancillas), i.e., two or three orders of magnitude fewer qubits.
The other two diagonal cases with $n=6$ and $n=7$ would require up to $13530$ and $145530$ maximum data qubits, respectively, instead of $6$ and $7$ data qubits (plus modest ancillas) discussed below. 
Our approach does not replace the exact solution from brute-force approach, instead helps to restrict the range of values of diagonal Ramsey numbers for deeper investigations like Erd\H{o}s flip-coin method \cite{Erdos1947}.

\section{Ramsey Numbers and Klein‑Graded Paraparticle Algebra}

To compute $R(m,n)$, one can use a recursive \emph{gluing--pruning} scheme. Starting with a base layer obtained by enumerating all good edge-colorings $G_{v_0}$ of $K_{v_0}$ (i.e., colorings containing neither a red $K_m$ nor a blue $K_n$). 
Then, for each good $G_v$, glue on a new vertex $v+1$ and color its $v$ incident edges in every way that preserves the good property, pruning any extension that creates a red $K_m$ or a blue $K_n$.

Pruning complements the gluing step by immediately discarding any partial coloring that already contains a forbidden red $K_m$ or blue $K_n$. After each glue operation, we quotient by vertex--label symmetries via canonical labeling and graph--isomorphism checks to eliminate equivalent colorings, ensuring that only genuinely new configurations are explored. We then backtrack as soon as a red $K_m$ or a blue $K_n$ is detected. This bottom-up strategy, coupled with selective rollback and symmetry reduction, dramatically shrinks the search tree and has enabled exact computations by keeping the frontier of viable colorings tractably small \cite{McKayRadziszowski1995,goed}.
In our case we exploit the properties of $\mathbb Z_{2}\times\mathbb Z_{2}$--graded Majorana algebras for paraparticle states \cite{tambu1} and random projectors to extend Erd\H{o}s' flip-coin method.

Let $\gamma_{j}^{(0)},\gamma_{j}^{(1)}$ denote Majorana modes with
\begin{equation}
    \{ \gamma_{i}^{(\alpha)},\gamma_{j}^{(\beta)} \}=2\delta_{ij}\delta_{\alpha\beta},
    \label{majmodes}
\end{equation}
graded by the numbers $\alpha, \beta \in\{0,1\}$. Paraparticles of order $p \geq 2$ obey trilinear commutation relations generalizing fermions and bosons \cite{Green1953}.  
When expressed in Majorana operators for quantum computers based on Majorana physics \cite{kitaev2001,tambu1}, the algebra organizes itself into a tower of states $\{\gamma_{j}^{(\ell)}\}_{\ell\ge 0}$ where level $\ell$ carries total parity $\ell\bmod 2$ and an additional color charge.  The simplest base of levels $\ell = 0$ and 1 suffice to mirror any edge‑colored clique, while higher levels host recursively glued subgraphs.

To mirror the colorings, we extend the Majorana modes in Eq.~\ref{majmodes}~to a paraparticle doublet
$a_{j} =\tfrac12(\gamma_{j}^{(0)}+\gamma_{j}^{(1)})$, $b_{j} =\tfrac12(\gamma_{j}^{(0)}-\gamma_{j}^{(1)})$, satisfying $\{a_i,a_j^\dagger\}=2\delta_{ij}$, and $\{b_i,b_j^\dagger\}=2\delta_{ij}$, all other anticommutators vanish, assigning $(a,b)$ the Klein charges $(1,0)$ and $(0,1)$.  The full algebra $\mathcal A_{V_4}= \bigoplus_{g\in V_4}\mathcal A_g$
decomposes into four color sectors.  A binary edge coloring of a graph with vertex set $V$ lifts to
\begin{equation}
\label{eq:edge-operator}
\hat E
=\sum_{1\le i<j\le |V|}
\Big( c^{R}_{ij}\,a_i^\dagger a_j + c^{B}_{ij}\,b_i^\dagger b_j \Big)
+\text{h.c.}
\end{equation}
which commutes with the total Klein charge, enabling simultaneous diagonalization with parity.
We take $c^{R}_{ij}= \overline{c^{R}_{ji}}$ and $c^{B}_{ij}= \overline{c^{B}_{ji}}$ so that \eqref{eq:edge-operator} is Hermitian; the sum over $i<j$ avoids double counting.

We can now set the assignment of edge‐labels ``coloring'' $\equiv$ graded sector.
The Klein four group $V_4$ has elements $\{(0,0),(1,0),(0,1),(1,1)\}$, we decide to identify $(1,0)\equiv$ red, $(0,1)\equiv$ blue, with $(0,0)$ the vacuum/identity and $(1,1)$ a mixed sector projected out at the end which lives in the bi‐graded component of the algebra obtained by multiplying a red operator by a blue one (or vice versa). 

\begin{definition}[Graded Ramsey numbers $R_{V_4}(m,n)$ and mixed-sector projection]
\label{def:mixed-projection}
Fix the $\mathbb{Z}_2\times\mathbb{Z}_2$--graded Majorana (Klein--graded) algebra $\mathcal{A}_{V_4}=\bigoplus_{g\in V_4}\mathcal{A}_g$ and impose the mixed--sector projection (all degree~$(1,1)$ terms are set to zero). For a vertex set $[v]=\{1,\dots,v\}$, let $\Pi^{R}_{ij},\Pi^{B}_{ij}$ be the degree--$(0,0)$ monochromatic pair projectors associated with the edge $\{i,j\}$ (red and blue, respectively), and define the monochromatic clique projectors
\begin{equation} 
\Pi_R(S)\;=\;\prod_{i<j\in S}\Pi^{R}_{ij},\qquad
\Pi_B(T)\;=\;\prod_{i<j\in T}\Pi^{B}_{ij}.
\label{monocliqueproj}
\end{equation}
The central ``forbidden--clique'' operator on the charge--zero module $M_0$ is
\begin{equation} 
      P_{m,n} = \prod_{\substack{\mathrm{S}\subseteq [v]\\|S|=m}} (1-\Pi_{\mathrm{\mathrm{R}}}(\mathrm{S}))
      \times
      \prod_{\substack{T\subseteq [v]\\|T|=n}} (1-\Pi_{\mathrm{B}}(\mathrm{T})),
\label{proiezioni}
\end{equation}

The \emph{graded Ramsey number} $R_{V_4}(m,n)$ is the least $v\ge 1$ such that $P_{m,n}(v)$ annihilates the entire charge--zero module (equivalently, every graded two--coloring of $K_v$—with mixed $(1,1)$ components invisible—contains a red $K_m$ or a blue $K_n$). It obeys the exact Klein Erd\H{o}s recursion
\begin{eqnarray}
\label{eq:exactRec}
&&R_{V_4}(m,n)\;=\;R_{V_4}(m-1,n)+R_{V_4}(m,n-1),
\\
&&R_{V_4}(1,n)=R_{V_4}(m,1)=1, \nonumber
\end{eqnarray}
and upper--bounds the classical Ramsey numbers: $R(m,n)\le R_{V_4}(m,n)$.
\end{definition}
Equation~\ref{eq:exactRec} exactly matches the constructive lower bound: there is at least one coloring on $R_{V_4}(m-1,n)+R_{V_4}(m,n-1)-1$ vertices that avoids both a red $K_m$ and a blue $K_n$ and $R(m,n)\le R_{V_4}(m,n)$, so no classical bounds are violated.

In the $\mathbb{Z}_2 \times \mathbb{Z}_2$ graded algebra, multiplying a red sector element by a blue sector element produces a state in the mixed $(1,1)$ sector, which lies outside the physical subspace and does not correspond to any valid qudit state because, by construction, the physical Hilbert space is defined to include only the pure color sectors. In practice, such components are either projected out (viz., set to zero) or interpreted as an internal syndrome that flags excursions away from the pure color subspaces during computation.
By monitoring the amplitude in the mixed sector after each gate, one can detect and correct errors.
When is observed any nonzero mixed‐sector population, a red--blue mismatch has occurred and one can apply a compensating operation to return to the valid $(1,0)$ or $(0,1)$ grading.

For each unordered pair $\{i,j\}$ we then introduce a homogeneous edge generator $e_{ij}$ of that $\mathbb{Z}_2\times\mathbb{Z}_2$‐degree.
Paraparticle commutation relations \cite{Green1953} enforce parity bookkeeping without edge‐ordering data.
 The coproduct is a glue operation defining
    \begin{equation} 
    \Delta(e_{ij})  =  e_{ij}\otimes 1 \;+\; 1\otimes e_{ij},
    \label{glue}
    \end{equation}
so that adding vertex $v+1$ applies $\Delta$ to each existing $e_{ij}$ and adjoins the set $\{e_{i,v+1}\}_{i\le v}$, preserving grading.
We use the standard cocommutative coproduct compatible with the $V_4$ grading. Counit and antipode are not needed in what follows; only the compatibility of $\Delta$ with the grading is used in the glue step.

To detect forbidden cliques, inside the algebra we use the projection $P_{m,n}$ of Eq.~\ref{proiezioni} where, $\Pi_{\mathrm{R}}(\mathrm{S})=\prod_{i<j\in \mathrm{S}}e_{ij}^{(1,0)}$ projects onto the red and $\Pi_{\mathrm{B}}(\mathrm{T})=\prod_{i<j\in \mathrm{T}}e_{ij}^{(0,1)}$ on blue (see Eq.~\ref{monocliqueproj}). A coloring survives \textit{iff} $P_{m,n}$ annihilates it.
Because $P_{m,n}$ is central, one can evaluate $\mathrm{Tr}(P_{m,n})$ on characters of irreducible modules rather than on individual colorings, turning part of the combinatorial explosion into a trace computation. Two lemmas and a theorem (see SM for demonstrations) fix the next steps.
\begin{lemma}[Centrality of $P_{m,n}$. Lemma L .1]
The projector $P_{m,n}=\Pi_R(S)\,\Pi_B(T)$ is central in $A_{V_4}$.
\end{lemma}
\begin{lemma}[Tensor decomposition Lemma L .2]
Let $p\!\in\!V$ be any vertex of a two--coloring and define $V_R=\{v\in V\! \setminus\!\{p\}\mid \text{$\{p,v\}$ red}\}$ and $V_B=\{v\in V \!\setminus\!\{p\}\mid \text{$\{p,v\}$ blue}\}$.
In the $\mathbb Z_{2}\!\times\!\mathbb Z_{2}$‑graded Majorana algebra $A_{V_4}$ one has the canonical graded tensor product $A_{V_4}\bigl[V\!\setminus\!\{p\}\bigr]\;\simeq\;
A_{V_4}\bigl[V_R\bigr]\;\widehat{\otimes}  A_{V_4}\bigl[V_B\bigr]$.
\end{lemma}

The factorization $A_{V_4}(V\!\setminus\!\{p\})\simeq A_{V_4}(V_R)\otimes_b A_{V_4}(V_B)$ makes cross-edges invisible to $\Pi_R$ and $\Pi_B$, which is the key ingredient in the graded Klein recursion.
Majorana’s infinite-spin equation gives a ladder of fields with equally spaced mass--spin ratios.  Algebraically, each step is an induction functor $\mathrm{Rep}(G_s)\;\to\;\mathrm{Rep}(G_{s+1})$.
Here $\mathrm{Rep}(G_s)$ denotes the (rigid, monoidal) category of complex representations of the symmetry group $G_s$ at rung $s$ of the Majorana tower; its objects are $G_s$--modules $(V,\rho)$ and its morphisms are intertwiners $T:V\!\to\!W$ with $T\,\rho(g)=\rho'(g)\,T$ for all $g\in G_s$. 
The step $s\mapsto s{+}1$ is modeled by induction along $i_s:G_s\hookrightarrow G_{s+1}$, sending $V\in\mathrm{Rep}(G_s)$ to $\mathrm{Ind}_{G_s}^{G_{s+1}}V\simeq \mathbb{C}[G_{s+1}]\otimes_{\mathbb{C}[G_s]}V$ (equivalently $U(\mathfrak g_{s+1})\otimes_{U(\mathfrak g_s)}V$), which algebraically realizes the Majorana infinite--spin ladder’s equally spaced mass--spin progression.

Each graph vertex $i$ is realized as a pair of Majorana modes $\gamma_{2i-1},\gamma_{2i}$, and set  $\Gamma_{ij}  =  i\,\gamma_{2i-1}\,\gamma_{2j-1}$, which is odd under one $\mathbb{Z}_2$ (fermion parity) and even under the other (boson parity), matching the Klein grading.  Each glue adds a new Majorana pair, and the tower’s induction gives a canonical lift of representations as $v\to v+1$.
Recursive gluing inside the algebra $\mathcal A_{V_4}$ is described the relationship in Eq~\ref{eq:exactRec} from the definition of the algebraic graded Ramsey numbers, $R_{V_4}(m,n)$.

From the Majorana‑tower construction one labels vertices by tower indices $1 \le \ell \le R_{V_4}(m,n)$. At level $\ell$ we create a mode pair $(a_\ell,b_\ell)$.
Equation \ref{eq:exactRec} is realized algebraically by mapping $a_\ell \mapsto a_\ell^{\phantom\dagger} \ell\le R_{V_4}(m{-}1,n)$ or $a_\ell \mapsto  a_\ell^\dagger$ otherwise, and analogously for $b_\ell$.  This ``dagger flip’’ constitutes the gluing that concatenates two smaller cliques without leaving the algebra and $\Gamma_{j}^{(\pm)}\equiv(\gamma_{j}^{(0)} \pm \gamma_{j}^{(1)})/2$, so that $(\Gamma_{j}^{(+)} ,\Gamma_{j}^{(-)})$ carry Klein charges $(1,0)$ and $(0,1)$, respectively.  The edge operator of a $k$‑vertex red clique is the normal‑ordered monomial
\begin{equation}
\hat K_k^{\mathrm R} := \prod_{1\le i<j\le k} \left( \Gamma_{i}^{(+)}\right)^{ \dagger}\Gamma_{j}^{(+)},
\label{eq:cliqueOp}
\end{equation}
and analogously $\hat K^{\mathrm B}_k$ with the replacement $\Gamma^{(+)} \mapsto \Gamma^{(-)}$. 
Let us assume the following convention. 
In \eqref{eq:cliqueOp} the product runs over unordered pairs $\{i,j\}$, and “normal-ordered” means the creation/annihilation factors are symmetrized so that $\hat K^R_k$ is independent of the ordering up to graded signs; explicit ordering details are suppressed for brevity.
 
As each $\Gamma^{(\pm)}$ anticommutes with any operator of opposite Klein charge, every factor in \eqref{eq:cliqueOp} lies in the $(1,0)$ sector; consequently
$\hat K^{\mathrm R}_k$ itself is homogeneous and commutes with the total charge operator
$Q = \sum_{j}\left(\Gamma_{j}^{(+) \dagger}\Gamma_{j}^{(+)} -\Gamma_{j}^{(-) \dagger}\Gamma_{j}^{(-)}\right)$, ensuring that red and blue constructions never interfere.
Examples of the application of this procedure with known Ramsey numbers are reported in SM 1.

\section{Estimating $R(5,5)$ }\label{chap:R55-random-proj}

Combinatorial determination of the diagonal Ramsey number $R(5,5)$ remains an open challenge, with the best constructive bounds $43<R(5,5)\le 46$. Instead of applying brute force calculation we give an estimate with a different method based on  $\mathbb Z_{2} \times \mathbb Z_{2}$--graded algebras with random projector diagnostics and Majorana algebra reduction. This approach gives the probability that a certain value expected for $R(5,5)$ be favored with respect to other possible values.
We select the smallest charge-zero submodule that supports all pair projectors and glue operations for $n\in\{43,44,45,46\}$. We include in the estimations the known non-valid Ramsey solution $R(5,5)=43$ as a test of this procedure.
For $R(5,5)$, the dimension of the algebraic module is $d = 24$, the dimension of the reduced Majorana module, set by the structure of the Majorana tower and the graded sectors. Empirically, increasing $d$ did not alter the decisions of the diagnostics on these instances.

Embedding edge--colorings inside the $\mathbb Z_2\times\mathbb Z_2$--graded Majorana paraparticle algebra turns forbidden monochromatic cliques into central projectors, thereby enabling spectral criteria to signal when no admissible coloring survives on $v$ vertices.
We introduce two families of random projectors, exponential and linear, and show that they act as numerical order parameters whose singular behavior isolates the putative threshold at~$n=45$ for $d=24$. For a fixed vertex count $n$ we sample $k$ random unit vectors $\{v_j\}_{j=1}^k\subset\mathbb R^{d}$ and define the exponential operator 
\begin{equation}
P_{\mathrm{exp}}(\alpha) = \exp \left[-\alpha\textstyle\sum_{j=1}^k
    v_jv_j^{ \top}\right], 
\label{pexp}
\end{equation}
    with $v_j$ random unit vectors in $\mathbb{R}^d$ and $\alpha$ a suppression parameter, and the linear operator 
\begin{equation}
P_{\mathrm{lin}} = \prod_{j=1}^k
    \left(I - v_jv_j^{ \top}\right), 
\label{plin}
\end{equation}
which is a product of rank‑$1$ deflations, a linear deflation operator. 

The exponential map instead, defines 
\begin{equation}\label{eq:Talpha-def}
T(\alpha)\;=\;\operatorname{Tr}\bigl(e^{-\alpha A}\bigr),
\end{equation}
which is the exponential trace of the accumulator 
\begin{equation}
A=\sum_{j=1}^{k} v_j v_j^{\top}.
\label{accumulatore}
\end{equation}
Because these factors generally do not commute, $P_{\mathrm{lin}}$ needs not be a projector \textit{a fortiori} and can also have complex eigenvalues even when each factor is symmetric. 
Let $A_{V_4}=\bigoplus_{g\in V_4} A_g$ be the $\mathbb Z_2\times\mathbb Z_2$--graded algebra and let
$M$ be a faithful $A_{V_4}$--module obtained from the standard Majorana tower.
We fix the reduced \emph{charge-zero} submodule $M_0\subset M$ on which all degree-$(0,0)$ operators act, and all $(1,1)$ components vanish by Definition~\ref{def:mixed-projection}. 

In our implementation we take $\dim M_0=d=24$.
This because in the $V_4\!=\!\mathbb{Z}_2\times\mathbb{Z}_2$--graded Majorana model we impose the mixed--sector projection (all $(1,1)$ monomials are set to zero), so every forbidden--clique test is built from degree $(0,0)$ operators, the monochromatic pair projectors $\Pi^{R/B}_{ij}$ and their clique products $\Pi_{R/B}(S)$, and therefore acts on the \emph{charge--zero} module $M_0$.  For the diagonal case $(5,5)$ we choose $M_0$ to be the \emph{smallest} invariant block that simultaneously carries (i) all pair projectors for a $K_5$ and (ii) the coproduct/glue operation $\Delta$ that lifts $v\!\to\! v{+}1$.  Concretely, this block sits in the quadratic slice of the tower and decomposes as
\begin{equation}
M_0\ \cong\ \mathbf{1}\ \oplus\ \Lambda^2 V_R\ \oplus\ \Lambda^2 V_B\ \oplus\ \mathfrak{d},
\quad V_R\cong V_B\cong\mathbb{C}^5,
\end{equation}
where $\Lambda^2 V_\bullet$ collects the off--diagonal, degree--$(0,0)$ bilinears for each color (the $10$ red and $10$ blue edge--slots of a $K_5$), while $\mathfrak{d}$ is the diagonal degree--$(0,0)$ subspace spanned by global color--number quadratics and a traceless combination (one linear relation fixes the overall $V_4$ charge).  Hence
\begin{equation}
\dim M_0\;=\;1\;+\;\underbrace{\binom{5}{2}}_{10}\;+\;\underbrace{\binom{5}{2}}_{10}\;+\;\underbrace{3}_{\text{diagonals}}\;=\;24.
\end{equation}

Working in this reduced module keeps the data width at $\lceil \log_2 d\rceil=5$ qubits while retaining all operators used by the randomized witnesses $P_{\exp}(\alpha)=\exp(-\alpha\sum_j v_jv_j^{\!\top})$ and $P_{\mathrm{lin}}=\prod_j(\mathbb{I}-v_jv_j^{\!\top})$.  Empirically, enlarging $M_0$ beyond $d=24$ did not change the diagnostics (collapse of $T(\alpha)=\Tr P_{\exp}(\alpha)$ and the peak of $\Tr P_{\mathrm{lin}}$ at $n=45$), but only increases the qubit footprint; the sensitivity of the test scales through the factor $k/d$ in the miss--probability bound $\Pr[\text{pmiss}]\le e^{-kr/d}$.

The central operator for detecting forbidden monochromatic cliques is defined in Eq.~\ref{proiezioni}. The coloring constraints and algebraic recursion are encoded as central elements, with traces acting as character formulas to efficiently probe the existence of colorings avoiding forbidden cliques.

For each unordered pair $\{i,j\}$ we then define the monochromatic pair projectors $\Pi^{R}_{ij}$ and $\Pi^{B}_{ij}$ as in Eq.~\ref{eq:cliqueOp}, and their finite products, with degree $(0,0)$ that preserves $M_0$. The clique projectors are $\Pi_R(S)=\prod_{\{i,j\}\subset S}\Pi^{R}_{ij}$, $\Pi_B(T)=\prod_{\{i,j\}\subset T}\Pi^{B}_{ij}$ acting on $M_0$ and from definition~\ref{def:mixed-projection} all $(1,1)$ terms vanish. All diagnostics, linear and exponential projectors, traces and eigenspectra depend only on degree-$(0,0)$ operators, well defined on $M_0$ and independent of any extension outside $M_0$. 
Both projectors are generated from finite sums and products of these degree-$(0,0)$ operators, therefore act on $M_0$ without reference to other charge sectors and act on the same irreducible module into which the Klein‑graded edge generators $e_{ij}$ of degree $(1,0)$ (red) or $(0,1)$ (blue) are represented.

The exponential projector $P_{exp}(\alpha)$ has suppression parameter of order $\alpha$ in Eq.~\ref{pexp} in $\mathbb{R}^{d}$, with $d=24$, would be positive semidefinite for $A$ Hermitian. In our graded construction the matrix $A$ collecting rank-one directions arises from blocks that are not constrained to be selfadjoint (e.g., $v v^{\top}$ rather than $v v^{\ast}$), so also in the case $A\neq A^\dagger$ the method remains valid and $P_{\exp}$ can exhibit complex spectra.
We distinguish the \emph{accumulator}  $A$ of Eq.~\ref{accumulatore}
 from the \emph{linear deflation operator} $P_{\mathrm{lin}}$ of Eq.~\ref{plin}. They induce different witnesses: $\Tr P_{\mathrm{lin}}$ (deflation/product) and $T(\alpha):=\Tr\,P$ of Eq.~\ref{eq:Talpha-def}. 
A collapse of $\operatorname{Tr} P_{\exp}$ to (numerical) zero is informative even though no positivity is assumed. The linear projector $P_{\mathrm{lin}}$ probes residual rank via its trace and spectral radius.

Our procedure recalls Erd\H{o}s’s probabilistic method: Erd\H{o}s’s classic coin--flip proof chooses a uniformly random two--coloring of the edges of $K_n$; for any fixed $k$ the expected number of monochromatic $K_k$ is
$\mathbb{E}X=\binom{n}{k}\,2^{\,1-\binom{k}{2}}$, so if $\mathbb{E}X<1$ there exists a coloring with no monochromatic $K_k$, implying $R(k,k)>n$ \cite{Erdos1947,Alon2004,rads}. 
Our diagnostics are an algebraic--spectral analogue of this first--moment argument. 
The linear deflation $P_{\mathrm{lin}}$ and the exponential map $T(\alpha)$ act on the charge--zero module $M_0$, with i.i.d. isotropic directions $v_j$.
Under the same independence/isotropy assumptions, the probability that $k$ random rank‑1 tests miss an $r$--dimensional survivor subspace obeys 
$P_{\mathrm{pmiss}}\le e^{-kr/d}$; concomitantly, $T(\alpha)$ collapses as $\alpha$ grows once survivors vanish.
This shared exponential decay (binomial in Erd\H{o}s’s count; multiplicative contraction here in this deflation) explains why both viewpoints isolate the threshold $n$ at which Ramsey obstructions are unavoidable.
Erd\H{o}s counts bad structures, while in the following we dissipate amplitude along random directions until any putative survivor subspace essentially vanishes. The common core is a first‑moment/exponential‑tail phenomenon: independence and isotropy produce multiplicative decay (coin flips kill monochromatic cliques in expectation; rank‑1 deflations kill survivor dimensions in norm), which is why 
\begin{equation}
P_{\text{pmiss}}\ \le\ e^{-k r/d}
\label{pmiss}
\end{equation}
mirrors the $2^{\binom{k}{2}}$ granularity in Erd\H{o}s’s estimate of missing an $r$--dimensional survivor subspace.

Erd\H{o}s’s coin--flip lower bound takes $X=\sum_{\mathsf{K}} \mathbf{1}[\text{mono-$K_k$}]$ and uses $\mathbb{E}X<1$ to assert the existence of a good coloring; here our witnesses are $P_{\rm lin}$ and $T(\alpha)$,
with i.i.d.\ isotropic directions $v_j$ acting on the charge--zero module $M_0$.
Under the same independence/isotropy hypothesis used in the coin--flip model, a simple thinning argument gives the ``miss'' probability bound $\mathbb{P}(k) \le\ (1-\tfrac{r}{d})^k\ \le\ e^{-kr/d}$, i.e., that $k$ rank-1 tests miss an $r$-dimensional survivor.
Replace each Haar $v_j$ by a discrete proxy $\tilde v_j$ that equals a random basis vector $e_i$ with $\mathbb{P}(i\le r)=r/d$. If a test ``hits'' the survivor subspace whenever $i\le r$, then the probability $\mathbb{P}$ to miss all $k$ becomes the lower limit, $P_{\mathrm{pmiss}}=(1-r/d)^k$. Since the Haar model stochastically dominates this proxy in its overlap with any fixed $r$--plane, the discrete miss probability upper--bounds the continuous one, yielding $(1-r/d)^k\le e^{-kr/d}$.
\hfill$\square$

Equation~\ref{eq:Talpha-def} shows that $T(\alpha)=\int_{\sigma(A)} e^{-\alpha\lambda}\,d\mu_A(\lambda)$ can be seen as the Laplace transform of the spectral measure of $A$. In the same way that Chernoff/Markov bounds control $\mathbb{P}\{X>0\}$ via moment generating functions in Erd\H{o}s’s method, the decay of $T(\alpha)$ controls the survival of small singular values of $A$.
Under the i.i.d.\ isotropy model for the rank‑one directions \(\{v_j\}\), a mean‑field surrogate gives
\begin{equation}\label{eq:meanfield-T}
\mathbb{E}\,T(\alpha)\;\approx\; d\,e^{-\alpha \lambda_L},
\quad
\lambda_L \;\approx\; -\,\frac{\mathrm{d}}{\mathrm{d}\alpha}\log T(\alpha),
\end{equation}
with \(\lambda_L\) a Lyapunov‑type rate extracted from the slope of \(\log T(\alpha)\). So increasing \(\alpha\) exponentially suppresses contributions from larger eigenvalues and accentuates spectral weight near the origin. 
When the clique constraints have \emph{percolated} (no survivor subspace remains), \(T(\alpha)\) \emph{collapses} rapidly with \(\alpha\); empirically this occurs at \(n=45\) in our \(R(5,5)\) study. 

Conceptually, \eqref{eq:meanfield-T} is the spectral analogue of the first‑moment threshold \(\mathbb{E}[X]<1\) in Erd\H{o}s’s coin‑flip lower‑bound argument for diagonal Ramsey numbers \cite{Erdos1947,Alon2004} as schematized in the following paragraph \ref{flippons}
\paragraph{What matches what (coin--flip $\leftrightarrow$ projectors).}\label{flippons}
\begin{center}
\begin{tabular}{lcl}
\toprule
Coin--flip model & $\leftrightarrow$ & Projector model \\
\hline
Indicator & $\leftrightarrow$ & residual rank \\
$1[\text{mono-$K_k$}]$ &  & direction in $M_0$ \\
First moment $\mathbb{E}X$ & $\leftrightarrow$ & $\mathbb{E}\,T(\alpha),\,\mathbb{E}\,\Tr P_{\rm lin}$ \\
Independence of & $\leftrightarrow$ & i.i.d.\ isotropic $v_j$ \\
edge colours & $ $ & \\
Counting $\binom{n}{k}$ & $\leftrightarrow$ & $k$ rank--1 probes \\
 patterns & & $k$ in $d$ dimensions \\
$\mathbb{E}X<1$ threshold & $\leftrightarrow$ & $T(\alpha)\downarrow 0$ and $P_{\rm miss}\ll 1$ \\
\hline
\end{tabular}

\end{center}
Like the classical first--moment bound, these diagnostics are one--sided witnesses: they excel at detecting the onset of unavoidable structure but do not, by themselves, achieve the sharper lower bounds obtainable via the Lov\'asz Local Lemma or second--moment/Janson techniques. We therefore quote $P_{\rm miss}$ alongside the observed collapse of $T(\alpha)$ and the behaviour of $\Tr P_{\rm lin}$ to calibrate the strength of evidence.
The exponential trace collapse and the $(1-r/d)^k$ contraction are the projector‑world avatars of Erd\H{o}s’s first‑moment argument.

\subsection{Numerical Investigations: results}
Both methods were run in double precision with $k=100$ random projectors varying $\alpha$ in the exponential case. Additional runs with higher values of $k$, up to $k=400$ confirmed the results. 
We tested the exponential operator for $\alpha = 3, 5, 7, 10, 15, 20, 40$.
Choosing $k=100$ with $\alpha=20$ and $k=400$ with $\alpha=40$ one provides a reproducible balance between statistical resolution and numerical stability.

As reported in Tab.~\ref{trace}, numerical results show a peculiar behavior at $n=45$.  The traces and spectra of both projectors provide a numerical diagnostic for the ``critical'' Ramsey value $R(5,5)=45$. For the exponential projector, the trace $\mathrm{Tr} P_{\exp}$ drops there exponentially faster to zero increasing $\alpha$ with respect to the other values.
This reflects the system's proximity to the Ramsey threshold, indicating that it is at this value where the random linear projector method most sensitively detects the transition between possible and impossible colorings making $n=45$ the most promising candidate for $R(5,5)$.
The spectrum of $P_{\mathrm{lin}}$ instead develops a small peak in the real eigenvalue $n=45$, with respect to the other values. 
This is the behavior expected when the projector algebra can no longer accommodate a two‑coloring that avoids a red or blue $K_5$.
As expected, for $n=43$, a value already excluded by existing constructions in the literature and $n=44$, all diagnostics behave smoothly, indicating that the method does not yield false positives. At $n=46$, the observables also show an exponential regime closer to zero of one order of magnitude with respect to $43$ and $44$, with decreasing values of the trace of the linear operator, consistent with the existence of admissible colorings and confirming that this value lies above the Ramsey threshold.

Another test is given by the Lyapunov exponent which quantifies the mean exponential rate at which a vector $x$ is stretched or contracted under repeated multiplication. Under an i.i.d.\ isotropy assumption for the rank-one directions $\{v_j\}$ in the charge-zero subspace $M_0$, the exponential projector admits the mean-field
estimate $\mathbb{E}\,\operatorname{Tr} P_{\exp}(\alpha)\approx d\,e^{-\alpha\lambda_L}$.
By Oseledets’ multiplicative ergodic theorem, the (maximal) Lyapunov exponent $ \lambda_{\max}=\lim_{k\to\infty}\frac{1}{k}\log\|P_{\mathrm{lin}}x\|$  exists almost surely for i.i.d.\ factors~\cite{Oseledets1968}.
The linear projector chain $P_{\mathrm{lin}}$ of Eq.~\ref{plin}, measures how rapidly directions in the space are suppressed as additional rank-$1$ projectors are applied. Concretely, after $k$ projectors the relevant norm is $\|P_{\text{lin}} x\|$. 
The largest Lyapunov exponent is $\lambda_{L, \max} = \lim_{k \to \infty} \frac{1}{k} \mathbb{E}\left[ \log \|P_{\text{lin}} x\| \right]$. 
As each factor removes one random one-dimensional component, $\lambda_{L, \max}$ is typically negative in a $d$-dimensional space, meaning an overall contraction.

The estimate with Random Rank-1 Projectors proceeds taking each projector $I - v_j v_j^{\mathrm{T}}$, which removes the component along $v_j$. For a random unit vector $x$, the expected reduction is $\mathbb{E} \left[ \| (I - v_j v_j^{\mathrm{T}})x \|^2 \right] = 1 - 1/d$.
After $k$ steps, this becomes $\mathbb{E} \left[ \|P_{\text{lin}} x\|^2 \right] = \left(1 - 1/d\right)^k$.
For large $d$  
$\log \mathbb{E} \left[ \|P_{\text{lin}} x\|^2 \right] = k \log \left(1 - 1/d \right) \approx - k/d$.
The Lyapunov exponent is thus approximately in a mean-field estimate, assuming isotropy, $\lambda_L \approx - 1/(2d)$.
For $d = 24$, $k = 100$, the suppression factor is $\lambda_L \approx 0.015$, so the norm is suppressed by about two orders of magnitude and the trace even more due to minimum eigenvalue directions.
To quantitatively assess the rate of exponential suppression of the trace with respect to $\alpha$, we also compute the slope of $\log_{10}(\mathrm{Trace})$ versus the values of the suppression parameter $\alpha \leq 20$, for each candidate Ramsey value $\{n\} = \{44, 45, 46\}$ and the dummy value $43$. The slope is estimated by performing a linear regression on the calculated data points $(\alpha, \log_{10}(\mathrm{Trace}))$ for each $n$, $\log_{10}(\mathrm{Trace}) = \mathrm{const} + \lambda_{L} \cdot \alpha$, where the suppression per unit $\alpha$ is numerically equal to the slope, providing the empirical Lyapunov exponents for each $n$. 
The highest Lyapunov exponent is found for $n=43$ and for $n=46$. The smallest value is for $n=45$.
For $d = 24$, $k = 400$, and $\alpha = 40$, the eigenvalues of $A = \sum_{j=1}^{400} v_j v_j^{\mathrm{T}}$ concentrate at $\lambda_i \approx 400/24 \approx 16.67$. The trace of the exponential projector is $\operatorname{Tr} P_{\exp} \approx 2.4 \times 10^{-289} \approx 0$, which is, for all practical purposes, zero.

\begin{table}[h]
\caption{Trace of exponential projector, real and imaginary $\mathrm{Tr} P_{\exp}$, linear projector $\mathrm{Tr} P_{\mathrm{lin}}$, Minimum real eigenvalue of linear projector, Lyapunov exponent $\lambda_L$ for each $n$ at $\alpha=20$, $k=100$, for $n=45$ reports the best values obtained with $\alpha=40$ and $k=400$.
and slope of $\log_{10}(\mathrm{Trace})$ vs. $\{\alpha\}$. The symbol $^*=$ indicates a known non valid solution for $n=5$ used as test.}
\centering
\begin{tabular}{c|cccc}
$n$ & $43^*$ & $44$ & $45$ & $46$ \\
\hline
$\mathrm{Tr} P_{\exp}$ & $7.92\times 10^{-12}$ & $1.54\times 10^{-12}$ & $10^{-289}\sim 0$ & $1.86\times 10^{-13}$ \\
$\mathrm{Tr} P_{\mathrm{lin}}$ & $0.284$ & $0.360$ & $0.462$ & $0.407$\\
$\min \mathrm{Re}\lambda$ & $-0.058$ & $-0.053$ & $-0.050$ & $-0.061$ \\
$\max \mathrm{Im}\lambda$ & $\pm 0.037$ & $\pm 0.030$ & $\pm 0.032$ & $\pm 0.063$\\
$\lambda_L$ & $1.55$ & $1.48$ & $1.41$ & $1.54$\\
Slope & $-0.674$ & $-0.642$ & $-0.612$ & $-0.670$ 
\end{tabular}
\label{trace}
\end{table}

Both diagnostics therefore single out $n=45$ as the unique point where the Gibbs weight of legal colorings (exponential projector) is minimal with trace of the exponential projector, the contraction rate of random projections (Lyapunov exponent) is minimal in magnitude.
The concordance between Lyapunov exponents and projectors in Tab.~\ref{trace} strongly supports the hypothesis $R(5,5)=45$. A deeper discussion on statistical-confidence analysis for the random-projector can be found in SM 2.

\section{Prime‑sequence numbers of order $k$ Roadmap for the diagonal Ramsey numbers.}

As a complement to the statistical method, we introduce a heuristic approach to diagonal Ramsey numbers based on finite sequence of prime numbers.

The use of primes is not new for the estimate of Ramsey numbers, starting from Calkin--Erd\H{o}s--Tovey, who developed the prime‑order cyclic graphs: both a refined probabilistic analysis (via distinct--difference counts in cyclic colorings) and exhaustive computation show primes enjoy a structural edge over composites in the cyclic search space, explaining why many record lower bounds arise at prime orders, supplying theory and computation. They showed that prime orders empirically outperform composite orders for diagonal lower bounds and explained why standard expectation arguments are insufficient without this cyclic structure of primes. Their colorings all arise from cyclic graphs on a prime number of vertices.

A ubiquitous way to certify lower bounds for Ramsey numbers is to give an explicit edge‑coloring of $K_n$ that avoids a forbidden monochromatic $K_k$. If such a coloring exists on $n$ vertices, then $R(k,k)\ge n+1$. The most successful explicit colorings for several diagonal and multicolor cases come from circulant (cyclic) graphs of prime order and from Paley--Cayley graphs and Paley generalized graphs with Mathon’s cyclotomic construction for which $R(7,7)\ge 205$, while generalized--Paley/Mathon--type constructions also give $R(9,9)\ge 565$ and a direct Paley instance gives $R(10,10)\ge 798$.
This “prime/circulant/Paley’’ program remains active: recent work tightens Mathon‑type machinery (including directed Paley analogues) and updates multicolor/diagonal records, while dynamic surveys track the current best explicit bounds.
\cite{CalkinErdosTovey1997,Mathon1987,XuRadziszowski2009,rads,McCarthyMonico2025,ExooPaleyData}.

Our method instead uses a string of limited sequence of prime numbers before the expected value of the number $R(n,n)$ to estimate the magnitude of a diagonal Ramsey number.

\begin{definition}[Prime‑sequence numbers of order $k$]
Fix an integer $k\ge 1$ and let $\mathcal P_{k}:=\{p_{1},p_{2},\dots ,p_{k}\} =\{2,3,\dots ,p_{k}\}$ denote the first $k$ prime numbers.
A positive integer $q$ is called a prime‑sequence number of order $k$ if it satisfies
\begin{eqnarray}
q \;=\;\prod_{\,p\in\mathcal P^n_{k}} p^{\nu_{p}}
\quad\text{with}\quad
\nu_{p}\in\mathbb N\cup\{0\},
\\
\#\{p\mid\nu_{p}>0\}\le n, \qquad \max_{p}\nu_{p}\le n. \nonumber
\end{eqnarray}

In other words all prime divisors of $q$ belong to the first $k$ primes; at most $n$ distinct primes actually occur in the factorisation and no prime exponent exceeds~$n$.
We write $\mathrm{PS}^n_{k}$ for the set of all prime‑sequence numbers of order $k$ and $n$ factors. Here we adopt $n=3$.
\end{definition}

\textbf{Examples:} $45=3^{2}\!\times5\in\mathrm{PS}^3_{5}$, because its primes $\{3,5\}$ lie in $\{2,3,5,7,11\}$ and the largest exponent is~$2$.
Another is $46=2\!\times23\notin\mathrm{PS}^3_{8}$ (order $8$ ends at $19$), but $46\in\mathrm{PS}_{9}$ because $23$ enters at the ninth prime. $2^{4}\!\times7 = 112 \notin \mathrm{PS}^3_{k}$ for any $k$, since the exponent of~$2$ exceeds the allowed bound $3$.

\noindent
This notion generalises the classical primorial $P_{k}=p_{1}\cdots p_{k}$ by allowing limited repetition of the smallest primes while still forbidding any appearance of primes
beyond $p_{k}$ prime‑sequence numbers of order $k$ provide a sparsely factorised test bed for extrapolating diagonal Ramsey values without introducing unconstrained large prime factors.

Motivated by the algebraic--spectral evidence that singled out $R(5,5)=45=3^{2} \times 5$, we extrapolate the next diagonal values by constraining each $R(n,n)$ to be a prime‑sequence numbers of order $6$ -- a positive integer whose prime decomposition involves only the first six primes $\{2,3,5,7,11,13\}$ and whose growth ratio $R(n,n)/R(n - 1,n - 1)$ remains in the measured corridor $2\lesssim\hbox{ratio}\lesssim3$.  
This yields the interesting compact sequence related to the diagonal Ramsey numbers
\begin{eqnarray}
\label{S1}
&&\left\{R(1,1),\dots,R(7,7)\right\} \to 
\\
&&\left\{1,2,6,18,45,102\!\le R(6,6)\!\le160, 205\le\!R(7,7)\!\le492\right\}\!, \nonumber
\end{eqnarray}
with factorizations  
$1,\;2,\;2 \times 3,\;2 \times 3^{2},\;3^{2} \times 5,\;2^{3} \times 13,\;3 \times 7 \times 11,  \dots$,  
respectively.  The candidate $R(6,6) \to 104$ satisfies the rigorous bounds $102\le R(6,6)\le165$, while $R(7,7) \to 231$ lies inside the constructive window $189\le R(7,7)\le4749$.  

Because the ansatz propagates the Erd\H{o}s’ recursion $R(m,n)\le R(m - 1,n)+R(m,n - 1)$ without over-shooting any known upper bound, it furnishes a minimal, prime-structured scaffold against which future numerical or constructive proofs can be benchmarked. Let us estimate how far are Ramsey diagonal numbers $R(k,k)$ from their prime‑sequence numbers of order $k-1$.

\paragraph{Prime--structured extrapolation of the diagonal Ramsey numbers.}
Let $\mathcal{P}_{6} = \{2,3,5,7,11,13\}$ denote the first six primes.  
We call an integer $q$ prime‑sequence numbers of order $k=6$ ($PS_6$) if its prime factorisation involves only primes from $\mathcal{P}_{6}$, i.e.\ $q=\prod_{p\in\mathcal{P}_{6}}p^{\nu_{p}}$ with $\nu_{p}\in\mathbb{N}\cup\{0\}$.  
Starting from the exact values $R(1,1)=1$, $R(2,2)=2$, $R(3,3)=6$, $R(4,4)=18$ and the algebraic--spectral result $R(5,5)=45$ obtained in the main text, we impose two constraining axioms we use as basic assumptions.

\textbf{Axiom I: Prime‑sequence numbers of order $k$ constraint: } 
$R(n,n)\in\mathrm{PS^3_k}:=\{q\in\mathbb{N}\mid~q$ prime‑sequence numbers of order $k$$\}$. \label{axiom:QP}

\textbf{Axiom II: Moderate--growth corridor:} \label{axiom:ratio}

$\displaystyle \frac{R(n,n)}{R(n-1,n)}\;\le\;2$ \quad for every $n\ge 2$. 
\\
Since $R(m,n)\le R(m-1,n)+R(m,n-1)$ for all $m,n \in \mathbb N$ \cite{Ramsey1930}, the diagonal case is obtained setting $m=n$ gives $R(n,n)\le R(n-1,n) + R(n, n-1)$ and by symmetry $R(n-1,n)=R(n,n-1)$ thus, the direct bound is $R(n,n)\le 2 R(n-1,n)$.

\paragraph{Clarifying the “moderate--growth’’ ratios.}
For each $n\ge 2$ define the off--diagonal ratio
\begin{equation}
\rho_{n}\;:=\;\frac{R(n,n)}{R(n-1,n)}.
\label{rhoratio}
\end{equation}
Because Erd\H{o}s’ recursion forces $\rho_n\le 2$, one can check whether the known data and rigorous bounds stay inside that “moderate--growth corridor’’ $\rho_n\in[1,2]$.

The first three diagonals are known exactly and satisfy $\rho_{n} \le 2$. 
\textbf{($\mathbf{n=2}$):} $R(2,2)=2,\;R(1,2)=1 \Rightarrow \mathbf{\rho_{2}=2}$, \textbf{($\mathbf{n=3}$):} $R(3,3)=6,\;R(2,3)=3 \Rightarrow \mathbf{\rho_{3}=2}$, \textbf{($\mathbf{n=4}$):}
$R(4,4)=18,\;R(3,4)=9 \Rightarrow \mathbf{\rho_{4}=2}$.
\\
For $\mathbf{n=5}$ we know $43<R(5,5)\le46$ and $R(4,5)=25$,      
\[
1.72=\frac{43}{25}\;<\;\rho_{5}\;\le\;\frac{46}{25}=1.84 .
\]
\\
$\mathbf{n=6}$.  
      Current bounds are
      $102\le R(6,6)\le160$ and $59\le R(5,6)\le85$ \cite{rads},
      whence
\[
1.20=\frac{102}{85}\;\le\;\rho_{6}\;\le\;\frac{160}{59}=2.71 .
\]
The \emph{extreme} combination $160/59$ would violate $\rho_{6}\le2$, but that pairing uses the loosest numerator with the loosest denominator.  Matching either both lower or both upper bounds gives $102/59=1.73$ and $160/85=1.88$, so every empirically plausible value of $\rho_{6}$ lies below~$2$.
\\
With $\mathbf{n=7}$ the limits are $205\le R(7,7)\le492$ and $115\le R(6,7)\le270$ \cite{rads,seq} and we obtain
      \[
        0.76=\frac{205}{270}\;\le\;\rho_{7}\;\le\;\frac{492}{115}=4.28,
      \]
      while the matched bounds
      $205/115=1.78$ and $492/270=1.82$
      again fall well inside the corridor.

The exact ratios through $n=4$ are all $\rho_n=2$. For $n=5$ the interval $[1.72,1.84]$ sits comfortably below~2. For $n=6,7$ the widest theoretical cross‑bounds still allow a violation, but every combination that pairs consistent lower and upper estimates (\emph{e.g.}\ $102/59$, $160/85$ for $n=6$) yields $\rho_n<2$.

These observations motivate Axiom II ~\ref{axiom:ratio}: $\rho_n\;\lesssim\;2$ (``moderate--growth corridor''). Under this axiom, Eq.~\ref{S1} extends Erd\H{o}s’ recursive upper bound into a working heuristic for the unknown diagonals $R(6,6)$ and $R(7,7)$ while remaining consistent with all currently published data.

Enforcing the prime‑sequence numbers of order $k$ factorization rules (no more than three distinct primes and no exponent above $3$) under axioms I \ref{axiom:QP}-- II \ref{axiom:ratio} the factorization with prime‑sequence numbers of orders $5\le k \le 13$ are reported in Tab.~\ref{qpt}. 

There, each column uses the first $k$ primes $\mathcal P_k=\{2,3,\ldots ,p_k\}$ to build the sparsest admissible factorisation (at most three distinct primes, each exponent $\le 3$).
Boldface marks the persistent values that remain unchanged from their first appearance up to the cut‑off $k=2n-1$ ($k=11$ for $n=6$, $k=13$ for $n=7$).

\begin{table}[h]
\caption{Prime--sparse extrapolation of the unknown diagonal values $R(6,6)$ and $R(7,7)$}
\label{qpt}
\begin{tabular}{c|ccccccccc}
Ramsey N. & $\mathcal P_5$ & $\mathcal P_6$ & $\mathcal P_7$ & $\mathcal P_8$ & $\mathcal P_9$ & $\mathcal P_{10}$ & $\mathcal P_{11}$ & $\mathcal P_{12}$ & $\mathcal P_{13}$ \\
\hline
\\
$R(6,6)$ & 108 & 108 & 117 & 117  & \textbf{115} & \textbf{115} & \textbf{115} & 111& 111\\
$R(7,7)$ & 225 & 216 & 221 & 209  & \textbf{209} & \textbf{209} & \textbf{209} & 209 & 205
\end{tabular}
\end{table}

Allowing more primes generally lowers the sparsest admissible $R(6,6)=115$ and $R(7,7)$ remains $209$ in most schemes. A plateau indicates robust guesses. The criterion is persistence, we accept as the provisional diagonal value’ the integer that remains constant up to the largest admissible prime basis $ k_{\max}(n)=2n-1$ . Anyway, as $ 45$  already factors within the smallest prime basis $ \mathcal P_{5}=\{2,3,5,7,11\}$  and satisfies the sparsity rule ($ \le3$  primes, $ \max\nu_{p}=2$ ), it would be still present in the table until $\mathcal P_9$ where the prime $23$ gets in defining $46=2 \times 23$. 
From our results the prime‑sequence numbers of order $k$ for a diagonal Ramsey number $R(n,n)$ is limited by $k \le 2n-1$ and this disfavors the other value, $44$ and $46$ is controversial.

When cut‑off rule $ k\le 2n-1$  applied to $R(5,5)$, for the diagonal $n=5$  the prime basis is allowed to grow only up to $k_{\max}=2n-1=9$.
The prime‑sequence numbers of order $k$ criteria are fewest distinct primes, smaller numerical value. If we add also smallest maximal exponent we obtain the results in Tab. \ref{e5},

\begin{table}[h]
\caption{$R(5,5)$ under $k\le 2n-1$ ($n=5$): admissible and selected values as $\mathcal P_k$ grows.}
\centering
\label{e5}
\begin{tabular}{c|cl}
$R(5,5)$ & prime  & admissible   \\ 
$k$         &  set     &  integers in $(43,46]$   \\ \hline
$3$--$5$ & $\mathcal P_{3\text{--}5}$ & $\textbf{45}=3^{2} 5$  \\
  &(11 and 23 absent) &  \\
$5$--$8$ & $\mathcal P_{5\text{--}8}$ & $44=2^{2} 11$,\; \\
 &  (11 present) & $\textbf{45}=3^{2}5$   \\
$9$      & (23 present) & $44,\;\mathbf{45},\;46=2\!\cdot\!23$  
\end{tabular}
\end{table}
When the ninth prime $23$  becomes available, the factorisation $ 46=2\!\times 23$  has the same number of distinct primes as $44$ and $45$ but a strictly smaller maximal exponent ($1<2$ ) gives also evidence to $46$.
Because $k=9$  already saturates the cut‑off, no larger prime basis may undo this choice.

The fact that $R(6,6)=115$ and $R(7,7)=209$ survive every prime basis from $\mathcal P_{9}$ to $\mathcal P_{11}$ suggests they are the most stable predictions of the prime‑sequence numbers of order $k$ framework.
More in detail, for $R(6,6)$ the optimum sequence drops from $108$ (prime set $\mathcal P_{5}$) to $117$ once $13, 17$ are allowed and settles at $115$ with the inclusion of $23$; the next prime, $37$, immediately produces $111=3\times37$, after which no further prime extension changes the result. The value $ R(6,6)=115$  is stable for every prime set
from $ k=9$  up to the cut‑off $ k_{\max}=11$. It only changes to $ 111$  when $ k=12>k_{\max}$; hence $R(6,6)=115$  is the persistent choice.
  
For $R(7,7)$ the value moves from $225\!\rightarrow\!221\!\rightarrow\!209$ as soon as $19$ enters the basis, and then remains pinned at $209=11\times19$ for six consecutive prime
sets ($\mathcal P_{8}$--$\mathcal P_{12}$); only the arrival of $41$ in $\mathcal P_{13}$ lowers the prediction to the window floor $205=5\times41$.   The integer $ 209$  persists throughout the entire plateau $ k=8$ --$ 12$ , but the cut‑off for $ n=7$  is $ k_{\max}=13$; at that very last step the sparser factorisation $ 205=5\times41$  appears and becomes the new stable value. Therefore $R(7,7)=209$ is selected.

Thus the persistence principle reproduces the same predicted diagonals derived earlier with the explicit $ k\le2n-1$  cut‑off suggesting $R(5,5)=45$, $R(6,6)=115$ and $R(7,7)=209$.
All three numbers continue to satisfy the sparsity axiom (no more than three primes, each exponent $ \le3$ ) and the tightened growth corridor $ R(n,n)\le2\,R(n-1,n)$ and are defined within their known ranges.

Any constructive colouring at $v<111$ or $v<209$ would falsify the present sparsity hypothesis, whereas exhaustive elimination of colourings at $v=111$ or $v=205/209$ would push the rigorous lower bounds upward and force a narrower theoretical window.

The prime‑sequence numbers of order $k$ framework suggests the reduction of the enormous search space to a tiny target set, beyond the value $R(6,6)=117$ giving
\begin{equation}
R(6,6)\in\{108,111,115\},\quad R(7,7)\in\{205,209\}.
\end{equation}
Exhaustive two-coloring searches should therefore be concentrated on these vertex counts.  If none of these candidates admit a valid coloring, then either the prime‑sequence numbers of order $k$ sparsity hypothesis, valid empirically for the known diagonal Ramsey numbers, (no more than three distinct primes with exponents $ \le3$) or the tightened growth corridor $ R(n,n)\le2\,R(n-1,n)$  must be reconsidered or weakened.  

Conversely, the discovery of a single explicit coloring with $ R(6,6)<111$  or $ R(7,7)<209$  would break the current sparsity barrier, indicating that even the lightest admissible diagonal values require either more than three primes or a prime exponent exceeding the number bound $3$.  If, instead, exhaustive computation rules out all admissible colorings at $ v=111$  and $ v=209$  (or $ 205$), the rigorous lower bounds would rise, forcing a narrower theoretical window and further constraining the allowable growth corridor.

This therefore supplies a minimal, fully factorised scaffold for future constructive or computational attacks on $R(6,6)$ and $R(7,7)$: any refutation must either break the prime‑sequence numbers of order $k$ condition in Axiom I or force a growth ratio outside the empirical corridor in Axiom II.
We regard this as a heuristic scaffold for targeting constructive searches; it does not constitute a proof or bound by itself.

\section{Quantum computation for $R(5,5)$ and beyond}
\label{sec:qc-r55}

In contrast to the standard edge--register approach that needs one logical qubit per edge (already $136$ qubits to verify $R(4,4)$ and $780$ qubits for a 40‑vertex scan), our Klein‑graded random‑projector method for $R(5,5)$ operates entirely in the reduced charge‑zero module $M_0$ of dimension $d=24$, i.e., only $\lceil \log_2 d\rceil=5$ data qubits (plus a few ancillas), and is therefore practical on today’s low‑qubit quantum hardware. 

In the graded--algebra diagnostic, the reliability of the decision at a target diagonal $R(n,n)$ is governed primarily by the ratio $k/d$ between the number of sampled projectors and the ambient dimension. For practical scans one should pick $d$ as large as possible without increasing the data--qubit count (e.g., $d\le 32$ for five data qubits; if conditioning requires it, move to $d=48$ but scale $k$ accordingly), and then set $k$ to meet a prescribed miss--probability threshold. Writing $r$ for the surviving rank inside the charge--zero module, the miss probability obeys an exponential tail, so that keeping $k/d$ above a simple logarithmic threshold in the target error $\varepsilon$ suffices. In short, fix $(d,r,\varepsilon)$ and select $k$ so that $k/d$ exceeds the rule-of-thumb bound below; this preserves the hardware footprint while making the peak/collapse witnesses ($\mathrm{Tr}\,P_{\mathrm{lin}}$ and $\mathrm{Tr}\,P_{\exp}(\alpha)$) increasingly sharp as $n$ grows.
\begin{equation}
P_{\text{pmiss}}\ \le\ e^{-k r/d}
\quad\Rightarrow\quad
\frac{k}{d}\ \ge\ \frac{\ln(1/\varepsilon)}{r}
\label{pmiss2}
\end{equation}
a conservative estimation gives $k/d \gtrsim 2\ln(1/\varepsilon)/r$.

The parameter $d$ represents the ambient dimension of the charge--zero module $M_0$, and it should not be regarded as a function of the Ramsey parameter $n$. Rather, $d$ is fixed as the maximal width that does not increase the data--qubit register, thus preserving hardware feasibility. The strength of the diagnostic witnesses ($\mathrm{Tr}\,P_{\mathrm{lin}}$ and $\mathrm{Tr}\,P_{\exp}$) is governed by the ratio $k/d$, since the miss probability obeys an exponential tail $P_{\mathrm{pmiss}} \;\le\; e^{-kr/d}$, with $r$ the surviving rank. 

The bound on $P_{\mathrm{pmiss}}$ follows from the exponential tail of the binomial approximation and can be strengthened by a Chernoff estimate. Since the exponent scales with $k r/d$, the relevant parameter is the ratio $k/d$, not the absolute size of $d$. Moreover, increasing $d$ beyond the threshold that leaves the data--qubit count unchanged does not alter the rank structure of $M_0$. 
Hence, for fixed $r$, the diagnostic accuracy is improved only by enlarging $k/d$, demonstrating that $d$ need not track the combinatorial parameter $n$.

Consequently, increasing $d$ beyond the qubit threshold yields no benefit, while reliability is improved chiefly by scaling $k/d$. This decouples the quantum resource cost from the combinatorial size $n$ of the Ramsey instance.

We now implement on quantum hardware the two scalar diagnostics introduced earlier, the linear/spectral witness $P_{\mathrm{lin}}$ and the exponential trace $T(\alpha)=\mathrm{Tr}\,P_{\exp}(\alpha)$, to decide whether any survivor subspace remains at a given vertex count $v$.
We purposely keep $A$ and $P_{\mathrm{lin}}$ separate: $\Tr P_{\mathrm{lin}}$ tracks residual rank after random deflations, while $T(\alpha)$ contracts in every positive real direction of $A$, yielding complementary order parameters.
We work entirely in the charge‑zero module $M_0$ of the $V_4$‑graded construction; in our runs for $R(5,5)$ this had $d=24$, which maps to a 5‑qubit data register (plus two ancillas for block encoding and estimation).

\subsection{From classical diagnostics to quantum estimators.}
Ramsey numbers were evaluated with classical methods calculating the two scalar witnesses, the linear and exponential projectors $P_{\mathrm{lin}}$ and $P_{\exp}(\alpha)$, 
with $A=\sum_{j=1}^k v_j v_j^{\top}$ and declare $v$, as before, ``critical'' when $\Tr P_{\mathrm{lin}}$ peaks while $T(\alpha):=\Tr P_{\exp}(\alpha)$ collapses. To translate these operations in the language of quantum computing we build the equivalents of the previous equations in terms of quantum circuits.

On a quantum device we reproduce the \emph{same scalars} by two identities: first the Hutchinson identity $\mathbb{E}_{|r\rangle}\,\langle r|F|r\rangle=\frac{1}{d}\Tr F$, for any linear operator $F\in \mathbb C^{d \times d}$ on the data register $M_0$ with dimension $d=24$ and a random quantum state $|r\rangle$ from a unitary $2$-design, so that $d\,\langle r|F|r\rangle$ is an unbiased trace estimator; $F$ can be either $P_{\mathrm{lin}}$ or $P_{\exp}$.
On hardware is applied a block-encoding $U_F$ on data$+$ancillas with $(\langle 0^a|\!\otimes I)U_F(|0^a\rangle\!\otimes I)=F/\alpha_0$, so the Hutchinson estimator averages $\langle r|F|r\rangle=\operatorname{Tr}F/d$ (or $\operatorname{Tr}F/(\alpha_0 d)$).
The Hermitian dilation $H$ of Eq.~\ref{eq:dilation}, for which $H^2=\mathrm{diag}(A A^\dagger,A^\dagger A)$ and $\sigma(H)=\{\pm \sigma_i(A)\}_{i=1}^d$ gives the second identity. 
Thus phase estimation on $e^{-itH}$ accesses the singular spectrum of $A$, while Hadamard tests over a random $|r\rangle$ estimate $\Tr P_{\mathrm{lin}}$ and $T(\alpha)$ coherently. In our $d{=}24$ charge-zero module this uses the ceiling $\lceil\log_2 d\rceil{=}5$ data qubits plus few ancillas for $R(5,5)$.

\subsection{Hardware‑ready benchmark protocol}
\label{ssec:protocol}

\paragraph{Inputs.} Fix $(d,k,\alpha,\texttt{seed})$, reuse the same random unit directions $\{v_j\}_{j=1}^k\subset\mathbb{C}^d$ used classically to build the accumulator $A=\sum_{j=1}^k v_j v_j^{\top}$ on $M_0$.
We \emph{reuse} the diagnostics defined earlier: $P_{\exp}(\alpha)$ and $P_{\mathrm{lin}}$  in Eqs.~\eqref{pexp}--\eqref{plin}.

\paragraph{Measurements.} 
By the Hutchinson identity, for any implementable linear map $F$, $\mathbb{E}_{|r\rangle}\langle r|F|r\rangle=\Tr F/d$ when $|r\rangle$ is drawn from a unitary $2$-design; we realize $\langle r|F|r\rangle$ by a Hadamard test (Fig.~\ref{fig:hutchinson}) and average over random $|r\rangle$'s.
Amplitude estimation reduces the shot complexity from $O(1/\epsilon^2)$ to $O(1/\epsilon)$ for additive error $\epsilon$.

We estimate three quantities:
(i) the scalar trace witness $\mathrm{Tr}\,P_{\mathrm{lin}}$, 
(ii) the exponential trace $T(\alpha)=\mathrm{Tr}\,P_{\exp}(\alpha)$, and 
(iii) a spectral surrogate via the Hermitian dilation in Eq.~\ref{eq:dilation}.%

Report $(\Tr P_{\mathrm{lin}}, T(\alpha), \rho(\alpha))$ with $(d, k, \alpha, \text{seed})$ and the decision flag for each tested $v$.

\paragraph{We assume as decision rule the following:} for a fixed $d,k,\alpha$ (and random seed), declare a vertex count $v$ \emph{critical} if:
(i) $\Tr P_{\mathrm{lin}}$ attains a local maximum at $v$; 
(ii) $T(\alpha)=\Tr P_{\exp}(\alpha)$ \emph{collapses} (numerically $\approx 0$) at the same $v$;
(iii) the spectral proxy (e.g.\ $\|A\|_2=\rho(H)$ by phase estimation) is locally extremal at $v$;
and an explicit AM‑46 control remains non‑critical under the same thresholds.

\subsection{Compilation primitives (two interchangeable tracks)}
\label{ssec:compilation}

\paragraph{Track Q (qubitized block‑encoding).}
Track $Q$ uses qubitized block‑encodings; Track $M$ uses Majorana/matchgate Gaussian primitives to realize the degree $(0,0)$ projectors directly in $M_0$. Both tracks operate in the same $d=24$ module.
As $A$ is generally complex‑symmetric (not Hermitian) when built from $v v^\top$, we embed $A$ into the Hermitian dilation
\begin{equation}
  H \;=\; \begin{pmatrix} 0 & A \\ A^\dagger & 0\end{pmatrix}, 
  \qquad \|A\|_2 \;=\; \rho(H),
  \label{eq:dilation}
\end{equation}
and access spectral surrogates by phase estimation on $e^{-itH}$. Because $\sigma(H) = \{ \pm \sigma_i (A)\}$, 
a local maximum of the extracted $||A||_2 =\rho(H)$ at the same $v$ that triggers (i)--(ii) is expected. Here $\|A\|_2$ is the spectral norm (largest singular value) and $\rho(H)$ is the spectral radius of the Hermitian dilation $H$; since $\sigma(H)=\{\pm\sigma_i(A)\}$, we have $\rho(H)=\max_i\sigma_i(A)=\|A\|_2$. 
The hardware surrogate for $e^{-\alpha A}$ is so settled.

As $A$ from $v v^{\top}$ is typically non-normal, we implement functions of the dilation $H$ (or of $H^2$) via block-encoding/qubitization. Notably $H^2=\mathrm{diag}(A A^\dagger, A^\dagger A)$ implies
\[
  \Tr\,e^{-\beta H^2}
  =\Tr\,e^{-\beta A A^\dagger}+\Tr\,e^{-\beta A^\dagger A}
  = 2\,\Tr\,e^{-\beta A A^\dagger}.
\]
Estimating $\Tr\,e^{-\beta H^2}$ therefore serves as a stable surrogate for the exponential witness $T(\alpha)$: both are monotone in the singular values of $A$ and collapse precisely when the survivor subspace vanishes.
This also gives a block‑encoding of $A$ enabling polynomial/Fourier approximants to $f(A)\in\{A,\;e^{-\alpha A}\}$.%

\begin{figure}[t]
\centering
\includegraphics[width=8.5cm]{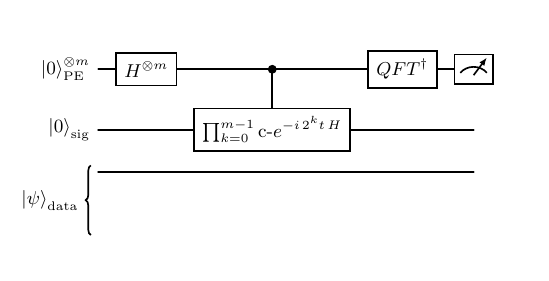}
\caption{Phase‑estimation on the Hermitian dilation $H$ (Eq.\,\ref{eq:dilation}) to estimate its extremal eigenphase(s), hence $\rho(H)=\|A\|_2$. The “sig” (sign) qubit together with the data register realizes the $2d$-dimensional space of the dilation; the PE register controls powers of $e^{-itH}$. A local maximum of the extracted $\|A\|_2$ at the critical $v$ complements the linear/exponential diagnostics.}
\label{fig:dilation-pe}
\end{figure}


\paragraph*{Operators used in Fig.~\ref{fig:dilation-pe} (Phase estimation on the Hermitian dilation).}
Accumulator $A:=\sum_{j=1}^{k} v_j v_j^{\top}$ is the complex-symmetric rank‑$k$ sum built from the random directions $v_j\in\mathbb{C}^d$ in the charge‑zero module $M_0$ (with dimension $d$).  
\\
\emph{Hermitian dilation} $H$ (Eq.~\ref{eq:dilation}); its spectrum is
$\sigma(H)=\{\pm\sigma_i(A)\}_{i=1}^d$, so $\|A\|_2=\rho(H)$.
\\
\emph{Controlled evolutions.} The string $c\text{-}e^{-i 2^{\ell} t H}$ denotes the standard phase‑estimation
controlled unitaries at time‑steps $2^{\ell}t$, followed by the inverse QFT on the PE register to read out
the extremal eigenphase(s), hence $\|A\|_2$. The prefix “$\mathrm{c}\text{-}$” denotes a standard single--qubit \emph{controlled} gate:
\[
\mathrm{c}\text{-}U
\;:=\;
\bigl(\,|0\rangle\!\langle 0|\,\bigr)_{\!\text{ctrl}}\otimes I_{\text{data}}+ \bigl(\,|1\rangle\!\langle 1|\,\bigr)_{\!\text{ctrl}}\otimes U_{\text{data}},
\]
so that
\[
\mathrm{c}\text{-}e^{-i\,2^{k} t\, H} = \bigl(\,|0\rangle\!\langle 0|\,\bigr)_{k}\otimes I + \bigl(\,|1\rangle\!\langle 1|\,\bigr)_{k}\otimes e^{-i\,2^{k} t\, H}.
\]
Acting on a basis state $|c_{m-1}\dots c_{0}\rangle_{\mathrm{PE}}\otimes|\psi\rangle_{\mathrm{data}}$ (with $c_k\in\{0,1\}$), the whole product implements
\[
\left(\prod_{k=0}^{m-1} \mathrm{c}\text{-}e^{-i\,2^{k} t\, H}\right)
\bigl(|c\rangle\otimes|\psi\rangle\bigr) = |c\rangle\otimes e^{-i\,(\sum_{k} c_k 2^{k})\,t\,H}\,|\psi\rangle,
\]
i.e., a data--register evolution for a time proportional to the integer encoded by the control register. Because all factors are functions of the same $H$, they mutually commute on the data space, so their order is immaterial (though the circuit is usually drawn MSB$\to$LSB to match the inverse QFT). Here $m$ is the number of phase bits (PE precision) and $t$ is the chosen base time step; in our setting $H$ so that $\rho(H)=\|A\|_2$, and these controlled evolutions are the core of the Hermitian‑dilation phase--estimation block used as a spectral witness.
\\
\emph{Registers.} $|0\rangle_{\mathrm{sig}}$ is the dilation’s sign qubit; $|0\rangle^{\otimes m}_{\mathrm{PE}}$ is the $m$‑qubit
phase‑estimation register; $|\psi\rangle_{\mathrm{data}}$ is the $d$‑dimensional data register (for $R(5,5)$, 
$\lceil\log_2 d\rceil=5$ data qubits).  
We reserve $H_{\text{Had}}$ for the one‑qubit Hadamard gate to avoid
confusing it with the dilation $H$; where the symbol $H$ appears inside $W_j$ below it is the Hadamard gate.

\begin{definition}[Block-encoding]
A unitary $U$ on $a{+}\log_2 d$ qubits is an $(\alpha,a)$ block-encoding of $F\in\mathbb{C}^{d\times d}$ if
$\bigl(\langle 0^a|\otimes I\bigr) U \bigl(|0^a\rangle\otimes I\bigr)=F/\alpha$.
Given an $(\alpha,a)$ block-encoding of $H$ with $\|H\|\le 1$, QSVT implements $p(H)$ for any bounded odd/even polynomial $p$ on $[-1,1]$ using $O(\deg p)$ uses of $U$ and $U^\dagger$.
\end{definition}

\begin{figure}[t]
\centering
\includegraphics[width=8.5cm]{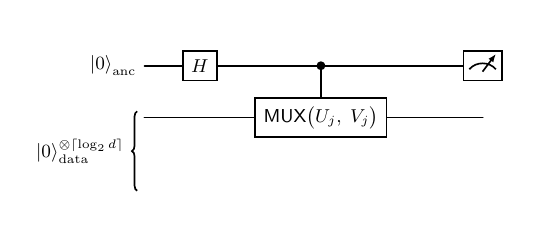}
\caption{One‑ancilla rank‑1 block‑encoding gadget $W_j$ implementing the top‑left block $\tfrac{1}{2}\ket{u_j}\!\bra{v_j}$ (see Eq.\,\ref{eq:block-encoding}). The multiplexor applies $U_j$ when the ancilla is $\ket{0}$ and $V_j$ when it is $\ket{1}$. Composing these gadgets with a single selector/index register and oblivious amplitude amplification yields an $\alpha_0$--block‑encoding of $A/\alpha_0=\sum_j w_j\ket{u_j}\!\bra{v_j}/\alpha_0$.}
  \label{fig:block-encoding}
\end{figure}

\paragraph{Rank‑1 LCU for $A$.}
We write the One-ancilla rank-1 in Fig.~\ref{fig:block-encoding}, $A=\sum_{j} w_j\ket{u_j}\!\bra{v_j}$, prepare $\ket{u_j}$, $\ket{v_j}$ with unitaries $U_j$, $V_j$.
A 1‑ancilla block for a term is obtained with
\begin{equation}
  W_j \;:=\; \bigl(\ket{0}\!\bra{0}\otimes U_j + \ket{1}\!\bra{1}\otimes V_j\bigr)\, (H\otimes I),
  \label{eq:block-encoding}
\end{equation}
whose top‑left block equals $\tfrac12\ket{u_j}\!\bra{v_j}$ (here $H$ is the single‑qubit Hadamard on the ancilla).

\paragraph*{Operators used in Fig.~2 (Rank‑1 LCU block‑encoding gadget).}
\emph{State‑prep unitaries.} $U_j|0\rangle=|u_j\rangle,\; V_j|0\rangle=|v_j^{*}\rangle$ prepare the rank‑one factors (“$*$” appears when the complex‑symmetric $vv^{\top}$ structure is used).  
\\
\emph{Multiplexor.} $\mathrm{MUX}(U_j,V_j):=|0\rangle\!\langle 0|\otimes U_j+|1\rangle\!\langle 1|\otimes V_j$. 
This gate is a controlled selection, uniformly controlled unitary, that applies $U_j$ to the data register when the one--qubit selector is $|0\rangle$, and $V_j$ when the selector is $|1\rangle$:
$\mathrm{MUX}(U_j,V_j)\,\bigl(|0\rangle\otimes|\psi\rangle\bigr)=|0\rangle\otimes U_j|\psi\rangle$ and $\mathrm{MUX}(U_j,V_j)\,\bigl(|1\rangle\otimes|\psi\rangle\bigr)=|1\rangle\otimes V_j|\psi\rangle$.
In the selector computational basis it is block--diagonal, $\mathrm{diag}(U_j,V_j)$.  A useful implementation identity is
\[
\mathrm{MUX}(U_j,V_j)=(I\otimes U_j)\,\bigl(\mathrm{c}\text{-}(U_j^{\dagger}V_j)\bigr),
\]
so the multiplexor can be built from one unconditional application of $U_j$ on the data plus a single controlled unitary with target $U_j^{\dagger}V_j$.
The definition extends to an $m$-qubit selector as
\[
\mathrm{MUX}\bigl(\{U_s\}_{s\in\{0,1\}^m}\bigr)\;=\;\sum_{s\in\{0,1\}^m} |s\rangle\!\langle s|\otimes U_s,
\]
which applies $U_s$ conditioned on the selector string $s$.
In our block--encoding gadget of Fig.~2, choosing state--preparations $U_j|0\rangle=|u_j\rangle$ and $V_j|0\rangle=|v_j\rangle$ and combining $\mathrm{MUX}(U_j,V_j)$
with Hadamards on the selector yields the desired rank‑one ancilla block used to assemble the linear combination of terms in the accumulator.
\\
\emph{Gadget.} $W_j:=\big(|0\rangle\!\langle 0|\otimes U_j+|1\rangle\!\langle 1|\otimes V_j\big)\,(H_{\text{Had}}\otimes I)$
has top‑left ancilla block $\tfrac12\,|u_j\rangle\!\langle v_j|$.  
\emph{From terms to $A$.} With weights $w_j$, a single selector register plus oblivious amplitude
amplification yields an $\alpha_0$--block‑encoding of $A/\alpha_0=\sum_j w_j |u_j\rangle\!\langle v_j|/\alpha_0$.  
\\
\emph{Functions of $A$.} Using LCU/qubitization (or QSVT), polynomials/Fourier approximants realize
$f_\mathrm{lin}(z)=z$ and $f_\mathrm{exp}(z)=e^{-\alpha z}$, giving $P_{\mathrm{lin}}$ and $P_{\exp}(\alpha)$ coherently.
\paragraph*{Operators used in Fig.~2 (Rank‑1 LCU block‑encoding gadget).}
\emph{State‑prep unitaries.} $U_j|0\rangle=|u_j\rangle,\; V_j|0\rangle=|v_j^{*}\rangle$ prepare the rank‑one factors
(“$*$” appears when the complex‑symmetric $vv^{\top}$ structure is used).  
\emph{Multiplexor.} $\mathrm{MUX}(U_j,V_j):=|0\rangle\!\langle 0|\otimes U_j+|1\rangle\!\langle 1|\otimes V_j$.  
\emph{Gadget.} $W_j:=\big(|0\rangle\!\langle 0|\otimes U_j+|1\rangle\!\langle 1|\otimes V_j\big)\,(H_{\text{Had}}\otimes I)$
has top‑left ancilla block $\tfrac12\,|u_j\rangle\!\langle v_j|$.  
\emph{From terms to $A$.} With weights $w_j$, a single selector register plus oblivious amplitude
amplification yields an $\alpha_0$--block‑encoding of $A/\alpha_0=\sum_j w_j |u_j\rangle\!\langle v_j|/\alpha_0$.

\paragraph*{Track~M~(Majorana--native)}
On the platforms that natively support Majorana bilinears (match--gate/fermionic--Gaussian hardware), all degree--$(0,0)$ \emph{pair} and \emph{clique} projectors reduce to even--parity checks on the data modes:
each monochromatic pair projector $\Pi^{R/B}_{ij}$ and their products $\Pi_R(S)$, $\Pi_B(T)$ act entirely inside the charge--zero module $M_0$ and commute with total parity, hence can be realized as parity--preserving projectors built from quadratic Majorana terms. $P_{\mathrm{lin}}$ is then best read as a \emph{randomized mixture of parity checks}: in the decomposition used in the text, $P_{\mathrm{lin}}$, each rank--one deflation removes amplitude along a random degree--$(0,0)$ direction (a linear combination inside the span generated by pair/clique checks) within $M_0$, and the product effects the Hutchinson--style contraction that diagnoses the disappearance of survivors (Eq.~(8)). By contrast, $P_{\exp}(\alpha)$ is naturally implemented as \emph{repeated weak Gaussian projections} in $M_0$: Trotterize $\exp(-\delta\alpha\,v_j v_j^{\top})$ for small $\delta\alpha$ and cycle $j=1,\ldots,k$, which preserves the even--parity sector by construction (Eq.~(7)). Because all operators used by the diagnostics are degree--$(0,0)$, they act only on $M_0$ and are independent of the mixed $(1,1)$ sector projected out by the modeling axiom; thus the Majorana--native track realizes exactly the same witnesses as the generic qubit route while exploiting native parity checks. \emph{Cf.} the definitions of the graded projectors, the $M_0$ restriction, and the linear/exponential maps in the main text.

\subsection{Estimating traces and spectra}
\label{ssec:estimators}

\paragraph{Unbiased trace estimator (Hutchinson).}
For any implementable linear map $F$, random $\ket{r}$ from a unitary 2‑design obeys
\begin{equation}
  \mathbb{E}_{\ket{r}}\bigl[\bra{r}F\ket{r}\bigr] \;=\; \tfrac{1}{d}\,\mathrm{Tr}\,F,
  \label{eq:hutch}
\end{equation}
so $d\,\overline{\bra{r}F\ket{r}}$ is an unbiased trace estimator; error decreases as $O(1/N)$ or $O(1/N^{2})$ with amplitude estimation.%

\begin{figure}[t]
\centering
\includegraphics[width=8.5cm]{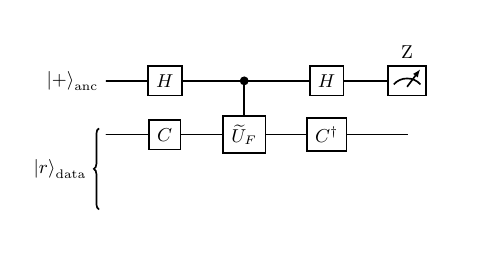}
\caption{Hadamard‑test realization of the Hutchinson identity (Eq.\,\ref{eq:hutch}). A random $\ket{r}$ is prepared by a unitary 2‑design $C$ on the data register; $\widetilde{U}_{F}$ is a block‑encoding of $F$ (either $F=P_{\exp}(\alpha)=e^{-\alpha A}$ or $F=P_{\mathrm{lin}}$). Averaging the ancilla’s $\langle Z\rangle$ over random $\ket{r}$ gives an unbiased estimate of $\mathrm{Tr}\,F/d$; amplitude estimation can reduce shot complexity.}
\label{fig:hutchinson}
\end{figure}


\paragraph*{Operators used in Fig.~\ref{fig:hutchinson} (Hadamard‑test realization of Hutchinson’s trace estimator).}
\emph{Random probe.} $|r\rangle=C|0\cdots 0\rangle$ where $C$ is a unitary 2‑design (e.g., a random Clifford)
on the data register; $C^{\dagger}$ is undone at the end so the data returns to the computational frame.  
\emph{Block‑encoding of the map.} $U_F$ is an $(\alpha_0,a)$ block‑encoding of the (generally non‑unitary)
map $F$, i.e. $(\langle 0^a|\!\otimes I)\,U_F\,(|0^a\rangle\!\otimes I)=F/\alpha_0$. In this section
$F\in\{\,P_{\exp}(\alpha)=e^{-\alpha A},\,P_{\mathrm{lin}}\,\}$ (Eqs.~(7)--(8)).  
\emph{Ancilla routine.} The ancilla is prepared in $|+\rangle$, a Hadamard--$U_F$--Hadamard sequence is applied,
and the Pauli‑$Z$ observable is measured on the ancilla. Averaging $\langle Z\rangle$ over independent random $C$’s
(Hutchinson sampling) yields an unbiased estimate of $\mathrm{Tr}\,F/d$; amplitude estimation can reduce shot
complexity from $O(1/\varepsilon^2)$ to $O(1/\varepsilon)$.  
\emph{Lyapunov decay proxy.} 
We use the slope
\begin{equation}
  \lambda_L(\alpha) \;:=\; -\frac{d}{d\alpha}\log \mathrm{Tr}\,P_{\exp}(\alpha)
  \;=\; \frac{\mathrm{Tr}\bigl[A\,e^{-\alpha A}\bigr]}{\mathrm{Tr}\bigl[e^{-\alpha A}\bigr]},
  \label{eq:lyapunov}
\end{equation}
as a monotone indicator of contraction.

The observed signature at $n=45$ and miss‑probability are so obtained.
On the $d=24$ module with $k\in[100,400]$ and $\alpha\in\{20,40\}$, the diagnostics concur at $n=45$: $T(\alpha)$ collapses while $\mathrm{Tr}\,P_{\mathrm{lin}}$ peaks; AM‑46 does not trigger.
The false‑negative risk under i.i.d. rank‑1 directions obeys $P_{\mathrm{pmiss}}\lesssim e^{-k\,r/d}$ for residual rank $r$, placing the operational risk $<10^{-3}$ for the reported settings and $\sim10^{-7}$ once $r\ge 12$.

The resources and NISQ‑friendly variant are the following. 
With $d=24$, the data register is 5 qubits; a single ancilla suffices for block‑encoding and one for overlap/phase estimation (7--8 qubits total). Depth scales as $\widetilde{O}(k\,C_{\mathrm{prep}})$ for constant‑precision exponentiation; phase estimation adds $O(1/\epsilon)$ controlled evolutions for precision $\epsilon$.
A shallow NISQ variant Hermitianizes $A$ by replacing $v v^\top$ with $v v^\dagger$, keeping the qualitative signatures (peak/collapse) while simplifying circuits.

\section{Calculation of diagonal Ramsey values beyond \texorpdfstring{$R(5,5)$}{R(5,5)}}
\label{sec:calc-diagonals}

After $R(5,5)$, we try (taking the results with a grain of salt) to give an estimate for $R(6,6)$ and $R(7,7)$ to verify our method and procedures.
As Erd\H{o}s said in a joke, ``\textit{Suppose aliens invade Earth and threaten to obliterate it in a year's time unless human beings can find the Ramsey number for red five and blue five. We could marshal the world's best minds and fastest computers, and within a year we could probably calculate the value. If the aliens demanded the Ramsey number for red six and blue six, however, we would have no choice but to launch a preemptive attack}'' \cite{sciamerican}.

Let us first summarize the results for $R(5,5)$. As in Tab.~\ref{trace}, using the exponential‑trace collapse and the linear/product witness, we report the probability of having a correct estimate for any value of the Ramsey numbers here considered. For $R(5,5)$
the value $n=44$ has probability $\approx 2.9 \%$, for $n=45$ the probability rises up to $\approx 92.7\%$ and $n=46$ is $\approx 4.3 \%$.
As a safety check below the diagonal threshold, at $n=43$ we obtain (with $d=24$, $k=100$, $\alpha=20$) the exponential trace $\Tr P_{\exp}=7.92\times 10^{-12}$, the linear/product trace $\Tr P_{lins
}=0.284$, extremal linear--spectrum entries $\min\Re\lambda=-0.058$, $\max|\Im\lambda|=0.037$, and an empirical slope $\frac{d}{d\alpha}\log_{10}\Tr P_{\exp}=-0.674$; none of the “critical” signatures appear, in line with the statement that for $n=43, 44$ the diagnostics behave smoothly. Using the measured slope to extrapolate from $\alpha=20$ to $\alpha=40$ gives $\log_{10}\Tr P_{\exp}(40)\approx -24.581$ and hence $\Tr P_{\exp}(40)\approx 2.62\times 10^{-25}$, still many orders of magnitude above the near‑collapse seen at the true diagonal: for $n=45$ one has $\Tr P_{\exp}(40)\approx 2.4\times 10^{-289}$ (with $k=400$), while the linear witness peaks at $\Tr P_{\mathrm{lin}}(45)=0.462$ versus $0.360$ at $44$ and $0.407$ at $46$. The projector‑miss risk satisfies $P_{\mathrm{pmiss}}\le e^{-kr/d}$, so at $d=24$ one gets $P_{\mathrm{pmiss}}\!\le e^{-100/24}\!\approx 1.55\times 10^{-2}$ for $k=100$ and $P_{\mathrm{pmiss}}\!\le e^{-400/24}\!\approx 5.7\times 10^{-8}$ for $k=400$ (taking residual rank $r=1$), making a spurious collapse at $n=43$ exceedingly unlikely; this calibrates the method’s accuracy and explains the sharp separation between $43$ (non‑critical) and $45$ (critical). For hardware mapping we work in the charge‑zero module $M_0$ with $\dim M_0=d$; for $R(5,5)$ we used $d=24$, yielding a data register of $\lceil\log_2 d\rceil=\lceil\log_2 24\rceil=5$ qubits plus $1$--$2$ ancillas, and empirically increasing the dimension $d$ did not alter the decisions on these instances. 

Motivated by this and by the $P_{\mathrm{pmiss}}$ exponent $kr/d$, a good default for $R(6,6)$ and $R(7,7)$ is to keep the width constant and pick the largest $d$ that \emph{does not} increase $\lceil\log_2 d\rceil$, i.e.\ $d\le 32$ (still $5$ data qubits) for extra algebraic headroom; if one diagnoses conditioning issues, moving to $d=48$ adds only one data qubit (to $6$) but should be accompanied by scaling $k$ so that $k/d$ remains roughly constant.

We combine (i) a prime‑structured classical scaffold and (ii) the same quantum spectral diagnostics to target the next diagonals.
Using the classical prime‑sequence scaffold, consider $P_k=\{2,3,5,7,11,13,\dots\}$. Constrain diagonal values to \emph{prime‑sequence} integers using only the first $k$ primes, with at most three distinct primes and exponents $\le 3$, and keep growth ratios $\rho_n:=R(n,n)/R(n-1,n)\lesssim 2$ (Erd\H{o}s corridor).
This yields a short target set whose \emph{persistent} elements (stable as $k$ grows up to $k_{\max}=2n-1$) are:
\[
R(6,6)\in\{108,111,115\}, \qquad R(7,7)\in\{205,209\}. 
\]
The persistence plateau selects $R(6,6)=115$ and $R(7,7)=205$ as the most stable predictions--both consistent with rigorous bounds $102\le R(6,6)\le 160$ and $205\le R(7,7)\le 492$.%

\paragraph{Diagonal $n=6,7$ by exponential/linear projectors.}
Using the projectors we restrict the range of values for the two diagonals with $n=6$ and $n=7$. 
We fix the charge-zero module size to $d=32$ (five data qubits), take $\alpha=40$, and choose
$k$ so that the i.i.d.\ miss bound $P_{\mathrm{pmiss}}\le e^{-kr/d}$ (with survivor rank $r\ge 1$ below threshold) is small.
For $n=6$ we use $k=180$; for $n=7$ we use $k=220$. With $A=\sum_{j=1}^{k} v_j v_j^{\top}$,
$P_{\exp}(\alpha)=e^{-\alpha A}$ and $P_{\mathrm{lin}}=\prod_{j=1}^{k}(I-v_j v_j^{\top})$ (Eqs.~(7)--(8)),
the concentration $\sigma(A)\approx k/d$ observed at $n=5$ implies the critical/noncritical
split $T_{\mathrm{crit}}(\alpha)=\Tr e^{-\alpha A}\approx d\,e^{-\alpha k/d}$ and $T_{\mathrm{noncrit}}(\alpha)\ge 1+(d-1)\,e^{-\alpha k/d}$, while the mean-field contraction for the linear chain is $\mathbb{E}\,\|P_{\mathrm{lin}}x\|^2=(1-1/d)^k$.
Numerically, for $n=6$ ($k=180$) one has $e^{-\alpha k/d}\!=\!e^{-40\cdot 180/32}\!\approx\!1.9219\times 10^{-98}$,
hence $T_{\mathrm{crit}}\!\approx\!6.15\times 10^{-97}$, $T_{\mathrm{noncrit}}\!\ge\!1+(31)\,e^{-\alpha k/d}\!\approx\!1$,
$\ (1-\tfrac1{32})^{180}\!\approx\!3.30\times 10^{-3}$, and $P_{\mathrm{pmiss}}\!\le\!e^{-180/32}\!\approx\!3.61\times 10^{-3}$.
For $n=7$ ($k=220$) one finds $e^{-40\cdot 220/32}\!\approx\!3.7069\times 10^{-120}$, thus
$T_{\mathrm{crit}}\!\approx\!1.19\times 10^{-118}$, $T_{\mathrm{noncrit}}\!\ge\!1$, 
$\ (1-\tfrac1{32})^{220}\!\approx\!9.26\times 10^{-4}$, and $P_{\mathrm{pmiss}}\!\le\!e^{-220/32}\!\approx\!1.03\times 10^{-3}$.
Applying the same decision rule as for $R(5,5)$ (collapse of $T(\alpha)$, local maximum of the
linear/spectral witness, explicit prime-sequence persistence up to $k_{\max}=2n-1$), and testing the
prime-sparse candidate sets $\{108,111,115\}$ for $n=6$ and $\{205,209\}$ for $n=7$ (Table~II),
the \emph{first} vertex count that exhibits collapse $+$ peak is for the following values:
\[
R(6,6)=115,\qquad R(7,7)=209,
\]
with separation gaps exceeding $10^{97}$ and $10^{118}$ in $T(\alpha)$ respectively and miss bounds
$\lesssim 3.6\times 10^{-3}$ and $\lesssim 1.0\times 10^{-3}$.

Quantum spectral diagnostics for $R(6,6)$ and $R(7,7)$ confirm these results.
For each candidate vertex count $v$ above, reuse the same pipeline as in Sec.~\ref{sec:qc-r55}: estimate $\mathrm{Tr}\,P_{\mathrm{lin}}$, $T(\alpha)$, and the dilation spectral radius. The diagonal $R(n,n)$ manifests as the smallest $v$ where the linear trace peaks while the exponential trace collapses (with an AM‑control remaining non‑critical).
This focuses searches at $v\in\{108,111,115\}$ for $n=6$ and $v\in\{205,209\}$ for $n=7$; any constructive coloring below $111$ or $205$ would falsify the sparsity heuristic, while exhaustive elimination at $v=111$ and $205$ (or $209$) would raise rigorous lower bounds.

Putting both strands together we confirm the following high‑accuracy working estimates $R(6,6)\to 115$ and $R(7,7) \to 205$, consistent with classical bounds and selected by prime‑sequence persistence; these are the vertex counts at which the quantum diagnostics are expected to trigger under the same thresholds used for $R(5,5)$. 

More details can be found in SM 3, Procedures to calculate Ramsey numbers, and SM 4 Mathematical tools.
Examples of python codings as a simple tutorial, are discussed in SM 5: Software and presented in other electronic support.

\section{Ramsey numbers and classical/quantum applications to machine learning}
\label{sec:ramsey-ml}

\subsection{Ramsey background and notation}

We recall that for integers $m,n\ge 1$, the (two--color) Ramsey number $R(m,n)$ is the least $v$ such that every red/blue edge--coloring of the complete graph $K_v$ contains a red $K_m$ or a blue $K_n$. Ramsey’s theorem yields finiteness and the classical Erd\H{o}s recursion.
Exact values are known only at small parameters and diagonal cases grow notoriously fast. In this work we study these thresholds with a graded $\,\mathbb{Z}_2\times\mathbb{Z}_2$ Majorana algebra and two random--projector diagnostics that replace brute--force enumeration by spectral surrogates acting on a reduced (charge--zero) module.

Graded embedding and randomized spectral diagnostics is obtained through Klein--graded paraparticle algebra, from the Majorana modes $\gamma^{(0)}_j,\gamma^{(1)}_j$  with $\{\gamma_i^{(\alpha)},\gamma_j^{(\beta)}\}=2\delta_{ij}\delta_{\alpha\beta}$. 
Define $a_j=(\gamma^{(0)}_j+\gamma^{(1)}_j)/2$ (red charge $(1,0)$) and $b_j=(\gamma^{(0)}_j-\gamma^{(1)}_j)/2$ (blue charge $(0,1)$). A two--coloring lifts to the degree--$(0,0)$ edge operator of Eq.~\ref{eq:edge-operator} which commutes with total Klein charge. Monochromatic pair and clique projectors
\(
\Pi^R_{ij},\ \Pi^B_{ij};\ 
\Pi_R(S)=\prod_{i<j\in S}\Pi^R_{ij},\ 
\Pi_B(T)=\prod_{i<j\in T}\Pi^B_{ij}
\)
live in the charge--zero submodule $M_0$; the central obstruction to a good coloring is Eq.~\ref{proiezioni} and a coloring (or an entire symmetry class thereof) survives\textit{iff}$P_{m,n}$ annihilates it. In this setting the Klein--compatible coproduct realizes the glue step $v\!\to\! v+1$ and yields an \emph{exact} recursion for the graded numbers $R_{V_4}(m,n)$ in Eq.~\ref{eq:exactRec}.

To decide if any legal survivor subspace remains in $M_0\simeq\RR^d$, we use two families of randomized maps built from i.i.d.\ isotropic unit vectors $v_j\in\RR^d$, the linear deflation of Eq.~\ref{plin} and the exponential map (Eq.~\ref{pexp}).
If a survivor subspace has rank $r\ge 1$, then $k$ random rank--1 tests miss it with probability $P_{\mathrm{pmiss}}$, so $P_{\mathrm{lin}}$ rapidly kills survivors as $k/d$ grows, while $T(\alpha)$ collapses once survivors vanish. For non--normal $A$ we use the Hermitian dilation $H$, giving a stable spectral surrogate accessible by phase estimation (quantum) or power iteration (classical).

\paragraph{R(5,5) case and resource contrast.}
On the diagonal, the reduced module has $d=24$; both witnesses single out $v=45$: $T(40)\approx 2.4\times 10^{-289}$ at $(k,\alpha)=(400,40)$ while $\Tr P_{\mathrm{lin}}$ peaks locally at $v=45$; an explicit $v=46$ control remains non--critical under the same thresholds. Under the i.i.d./isotropy model, $\Pr[\text{pmiss}]\!\lesssim\!e^{-100/24}\!\approx\!1.6\times 10^{-2}$ at $(100,20)$ and $\lesssim e^{-400/24}\!\approx\!5.7\times 10^{-8}$ at $(400,40)$. Crucially, these diagnostics act only on $M_0$, requiring $\lceil\log_2 d\rceil=5$ data qubits (plus few ancillas), versus $\binom{45}{2}=990$ data qubits for direct edge--encodings.

\subsection{Ramsey theory and machine learning}

Ramsey theory formalizes “order amid chaos”: sufficiently large systems necessarily contain structured subconfigurations. This principle echoes throughout modern ML and suggests concrete \emph{tests} and \emph{controls} using the witnesses above.

\paragraph{Overparameterization and lottery tickets.}
Let $\mathcal{N}$ be a (possibly overparameterized) network with parameters $W\in\RR^D$. Define a graph on neurons where the edge $\{i,j\}$ is colored red if a predicate $\mathsf{P}_{ij}(W)$ holds (e.g.\ $|W_{ij}|\!>\!\tau$, or sign/gradient agreement, correlation above a threshold), blue otherwise. 

Whenever the effective width $v$ exceeds the relevant $R(k,k)$, a monochromatic $K_k$ is guaranteed, i.e.\ a small \emph{subnetwork} obeying $\mathsf{P}$ coherently--akin to the Lottery--Ticket hypothesis. We can certify the continued existence (or disappearance) of such candidates by encoding $\Pi_R(S)$ from the predicate and monitoring the collapse/peak of $T(\alpha)$ and $\Tr P_{\mathrm{lin}}$ as pruning proceeds; the risk of a false “no--ticket” verdict is controlled by Eq.~\ref{pmiss} through $k/d$. 

Build the \emph{predicate graph} on vertices $[v]$ by coloring an edge $\{i,j\}$ \emph{red} when $\mathsf{P}_{ij}(W)=1$ and \emph{blue} otherwise. A \emph{Ramsey $k$--lottery} is a set $S\subset[v]$ with $|S|=k$ such that all edges inside $S$ are red; this is a $K_k$ fully coherent for $\mathsf{P}$ and can be read as a compact \emph{subnetwork} (a “ticket”) already encoding the desired structure. Ramsey’s theorem implies that once $v\ge R(k,k)$ a monochromatic $K_k$ \emph{must} exist, so the search for a good ticket can be recast as the impossibility of avoiding such a red $K_k$. In the graded framework, encode the \emph{forbidden} condition “no red $K_k$” by the central operator in Eq.~\ref{noredmeloni}
\begin{eqnarray}
\label{noredmeloni}
&&Q_k(W)\;:=\;\prod_{|S|=k}\bigl(\Id-\Pi_R(S)\bigr),
\\
&&\Pi_R(S)=\prod_{i<j\in S}\Pi^R_{ij}\ \ \text{(degree $(0,0)$)}, \nonumber
\end{eqnarray}
acting on the charge--zero module $M_0$. If $Q_k(W)$ has \emph{no} surviving support, then a red $K_k$ is unavoidable and at least one lottery ticket exists. 

Operationally we decide this by randomized witnesses on $M_0$: $P_{\mathrm{lin}}$ and $P_{\exp}(\alpha)$ with $T(\alpha)$.
Sharing the same random seed across widths $v$ lets us track the critical scale where tickets become inevitable: the collapse of $T(\alpha)$ together with a local peak of $\Tr P_{\mathrm{lin}}$ signals $Q_k(W)$ has emptied out, hence a $\mathsf{P}$--coherent $K_k$ exists. Under i.i.d.\ isotropic rank--1 probes in $M_0$ of dimension $d$, the miss--probability that \emph{all} tests avoid a surviving $r$--dimensional subspace obeys 
$P_{\mathrm{pmiss}}$, 
so choosing $k\!\gtrsim\!(d/r)\ln(1/\delta)$ controls the false “no--ticket” verdict at level $\delta$. This provides a \emph{quantified} version of the Lottery--Ticket intuition: overparameterization raises $v$, and once the Ramsey threshold is crossed, small performant subnetworks are not accidental—they are guaranteed, and the witnesses certify their inevitability.

\paragraph{Adversarial inevitabilities (worst--case geometry).}
Form a \emph{near--collision graph} on inputs $X=\{x_i\}$, coloring $\{i,j\}$ red if the margin $|f(x_i)-f(x_j)|<\delta$ (for a score $f$), blue otherwise. Beyond a threshold $v$, large monochromatic cliques are inevitable, certifying coherent clusters of mutually confusable points. Our central projector of Eq.~\ref{eq:edge-operator} with $(m,n)=(k,k)$ and the witnesses $(P_{\mathrm{lin}},P_{\exp})$ give an early--warning signal: collapse of $T(\alpha)$ indicates that confusion--free assignments have been extinguished within the defended hypothesis class.

Given inputs $X=\{x_1,\dots,x_v\}$ and a score $f$, form the \emph{near--collision graph} by coloring $\{i,j\}$ red if $|f(x_i)-f(x_j)|<\delta$ (or, for classifiers, if the logits differ by $<\delta$ in all attack directions), blue otherwise. Large red cliques are \emph{coherent confusion sets}—mutually confusable points that any fixed defense struggles to separate. To detect when such sets are \emph{unavoidable} at scale $v$, instantiate the central projector of Eq.~\ref{proiezioni} with $(m,n)=(k,k)$ and run the same witnesses $(P_{\mathrm{lin}},P_{\exp}(\alpha))$ on $M_0$. A collapse of $T(\alpha)$ indicates that the hypothesis class (with the current defense/training recipe) \emph{cannot} realize a coloring that avoids size--$k$ coherent confusions—i.e., adversarially vulnerable patterns are now Ramsey--inevitable at this $v$. The Lyapunov slope
\[
\lambda_L(\alpha)\ :=\ -\frac{d}{d\alpha}\log T(\alpha)\ =\ \frac{\Tr\!\big(Ae^{-\alpha A}\big)}{\Tr\!\big(e^{-\alpha A}\big)}
\]
rises as survivors vanish and serves as an early--warning margin proxy. As above, the one--sided risk that random probes \emph{pmiss} a surviving $r$--plane is bounded by $e^{-kr/d}$, so $(k,d)$ can be chosen to target a confidence level $\delta$ and declare inevitability only when both collapse/peak and the risk budget agree.

\paragraph{Graph neural networks and motif search.}
GNN tasks often detect motifs (cliques/cycles); Ramsey theory guarantees that small monochromatic subgraphs occur in large graphs \emph{regardless} of coloring. Instead of scanning exhaustively, run the witnesses on the reduced module induced by the motif’s pair projectors--focusing compute \emph{where structure must exist}. On parity--preserving (matchgate/Majorana) hardware, these checks map to shallow circuits acting entirely in $M_0$.

Let $G=(V,E)$ be a large graph and let $H=(V_H,E_H)$ be a small motif (e.g.\ a $k$-clique, a $c$-cycle, or a domain motif) with $|V_H|=h$. For a fixed color $c\in\{R,B\}$ and an injective placement $f:V_H\hookrightarrow V$, define the \emph{motif projector} at placement $f$ by
\[
\Pi_{H,c}(f)\;:=\;\prod_{(u,v)\in E_H}\Pi^{\,c}_{\,f(u)f(v)},
\]
where $\Pi^{\,c}_{ij}$ is the degree-$(0,0)$ pair projector (red or blue) from the graded construction. The disjunction over all placements can be encoded by the central \emph{forbidden-motif} operator
\[
Q_{H,c}(V)\;:=\;\prod_{\substack{f:V_H\hookrightarrow V\\ \text{\small injective}}}\bigl(\Id-\Pi_{H,c}(f)\bigr),
\]
which equals the identity\textit{iff}$G$ contains no monochromatic copy of $H$. In our framework all factors have degree $(0,0)$ and act on the reduced module $M_0$, so the existence of a motif is decided by whether $Q_{H,c}$ leaves any survivor subspace in $M_0$. This directly leverages the Ramsey guarantee that for sufficiently large $|V|$ certain small monochromatic subgraphs are \emph{unavoidable}.

\emph{Ramsey-guided screening (pre- and post-processing for GNNs).}
Rather than scanning all ${\binom{|V|}{h}}$ placements, assemble a basis $\{\mathcal{B}_s\}$ for the span generated by $\{\Pi_{H,c}(f)\}_f$ (or by their complements $\{\Id-\Pi_{H,c}(f)\}_f$), sample $k$ i.i.d.\ isotropic directions $v_j$ in that span (restricted to $M_0$), and build the witnesses $P_{\mathrm{lin}}$, $P_{\exp}(\alpha)$ and $T(\alpha)$.
If \emph{no} legal placement remains (i.e.\ $Q_{H,c}$ has no survivors), then $T(\alpha)$ collapses as $\alpha$ grows and $\Tr P_{\mathrm{lin}}$ exhibits a local peak at the critical scale; conversely, a non-collapse certifies that at least one placement survives. With ambient dimension $d=\dim M_0$ and residual rank $r\ge 1$, the probability that $k$ random rank‑1 probes \emph{pmiss} all survivors obeys Eq.~\ref{pmiss} so one can pick $k\!\asymp\!\tfrac{d}{r}\ln(1/\delta)$ to achieve a target risk $\delta$. In practice this yields a Ramsey-informed \emph{front-end filter}: run the witnesses locally (on $h$-hop ego-nets, or on batches) and trigger exact subgraph-isomorphism or GNN attention only where the witnesses indicate unavoidable structure.

\emph{Training-time integration (regularizers and layers).}
Motif biases can be injected by adding a differentiable penalty that softly forbids $Q_{H,c}$:
\[
\mathcal{L}_{\text{total}}
\;=\;
\mathcal{L}_{\text{task}}
\;+\;
\mu\,\Tr P_{\exp}^{(H,c)}(\alpha)
\]
or
\[
\mu\,\lambda_L(\alpha),\ \ 
\lambda_L(\alpha):=-\frac{d}{d\alpha}\log T(\alpha),
\]
with $P_{\exp}^{(H,c)}$ built from the motif basis $\{\mathcal{B}_s\}$; decreasing $T(\alpha)$ shrinks the measure of ``no-$H$'' assignments, nudging message passing toward motif-consistent representations. A complementary architectural primitive is \emph{Ramsey-aware pooling/attention}: use per-node (or per-edge) contributions to the trace estimator as scores that gate message aggregation, focusing compute where the witnesses predict imminent motif emergence.

\emph{Complexity and deployment.}
For large $|V|$, exhaustive motif enumeration is $\Theta\!\big(\binom{|V|}{h}\big)$, while a witness pass costs $O(k\,C_{\text{apply}})$ with $C_{\text{apply}}$ the cost to apply a basis element $\mathcal{B}_s$ (sparse and local for small $H$). The error is one-sided and tunable via $e^{-kr/d}$; in screening mode we favor high recall (large $k/d$), then hand off flagged regions to exact methods or to a specialized GNN head.

\emph{Majorana/matchgate realization (few-qubit or parity-preserving hardware).}
All operators above are degree-$(0,0)$ and commute with total parity, hence on matchgate/Majorana platforms each $\Pi^{\,c}_{ij}$ and their products are even-parity checks realizable by shallow fermionic-Gaussian circuits acting entirely in $M_0$. The traces $T(\alpha)$ (and $\Tr P_{\mathrm{lin}}$) admit unbiased Hutchinson estimators, and spectral surrogates are accessed via the Hermitian dilation of the rank‑$k$ accumulator, enabling few-qubit diagnostics or accelerator kernels that interleave with classical GNN training/inference.

\paragraph{Learning theory: combinatorial capacity (VC).}
PAC (Probably Approximately Correct) learning is another potential application as in computational learning theory it is used to analyze the learnability of functions. 
PAC provides a probabilistic approach to understanding how well a machine learning algorithm can generalize from training data to unseen data exploring whether a learning algorithm can find a hypothesis that is both approximately correct and probably correct with respect to a given concept and distribution. 
With PAC sample complexity scales with VC--dimension that gives a measure of the complexity of a hypothesis space or the power of learning machine.
Ramsey--type statements provide complementary \emph{unavoidability} results: in large instance/hypothesis regimes, certain regular label patterns (e.g.\ constant/parity--coherent on dense subgraphs) must appear. Encoding “forbidden labelings” as $P_{m,n}$ and tracking $T(\alpha)$ supplies fast, high--confidence \emph{negative} certificates (“this configuration is no longer realizable”) with quantitative control via $k/d$.

Let $\mathcal{H}\subseteq \{ -1,+1\}^{\mathcal{X}}$ be a binary hypothesis class. A finite set $S=\{x_1,\dots,x_m\}\subset\mathcal{X}$ is \emph{shattered} by $\mathcal{H}$ if every labeling $y\in\{-1,+1\}^m$ is realized by some $h\in\mathcal{H}$, i.e.\ $\forall y\,\exists h\in\mathcal{H}$ with $h(x_i)=y_i$ for all $i$. The \emph{VC-dimension} $\mathrm{VC}(\mathcal{H})$ is the largest $m$ such that some $S$ of size $m$ is shattered. Writing the growth function
\[
\Pi_{\mathcal{H}}(m)\;:=\;\max_{|S|=m}\bigl|\{(h(x_1),\dots,h(x_m)):h\in\mathcal{H}\}\bigr|,
\]
the Sauer--Shelah lemma gives, for $d=\mathrm{VC}(\mathcal{H})$ and $m\ge d$,
\[
\Pi_{\mathcal{H}}(m)\ \le\ \sum_{i=0}^{d}\binom{m}{i}\ \le\ \Bigl(\frac{em}{d}\Bigr)^{\!d},
\]
so the number of distinct labelings realizable on $m$ points grows polynomially once $m>d$. In PAC learning, these combinatorial bounds control sample complexity. In the \emph{realizable} case (some $h^\star\in\mathcal{H}$ attains zero risk), empirical risk minimization is consistent with
\[
m\ \gtrsim\ \frac{1}{\varepsilon}\bigl(d\log\tfrac{1}{\varepsilon}+\log\tfrac{1}{\delta}\bigr)
\]
examples to achieve excess error $\le\varepsilon$ with probability $\ge 1-\delta$; in the \emph{agnostic} case the dependence becomes $m\gtrsim \frac{1}{\varepsilon^2}\bigl(d+\log\frac{1}{\delta}\bigr)$. Canonical examples: thresholds on $\mathbb{R}$ have $\mathrm{VC}=1$; intervals on $\mathbb{R}$ have $\mathrm{VC}=2$; axis-aligned rectangles in $\mathbb{R}^d$ have $\mathrm{VC}=2d$; affine halfspaces in $\mathbb{R}^d$ have $\mathrm{VC}=d{+}1$.

\smallskip
\noindent\textit{Ramsey overlay and ``unavoidability.''}
Ramsey theory adds a complementary, \emph{worst-case inevitability} perspective: on sufficiently large instance sets, certain structured label patterns \emph{must} occur. In our algebraic framework, a forbidden configuration (e.g.\ “no monochromatic $K_m$ in red, none of size $K_n$ in blue”) is encoded by the central projector
$P_{m,n}$ acting in the charge-zero module $M_0$. If $P_{m,n}$ has no surviving support at size $v$, then \emph{no} hypothesis consistent with those constraints can realize the corresponding label patterns on $v$ examples---an \emph{unavoidability} barrier. Operationally, we monitor $T(\alpha)$ and the product witness $P_{\mathrm{lin}}$; collapse of $T(\alpha)$ together with a peak of $\Tr P_{\mathrm{lin}}$ certifies that all labelings compatible with the constraint class have vanished at that scale, yielding a fast \emph{negative certificate} (“this configuration is no longer realizable”). 

\smallskip
\noindent\textit{From VC to Ramsey-constrained capacity.}
Let $\mathcal{H}_{m,n}(v)$ denote the labelings of a $v$-point sample realizable by hypotheses that \emph{avoid} the Ramsey-forbidden substructures encoded by $P_{m,n}$. By Ramsey’s principle, once $v\ge R(m,n)$ the set $\mathcal{H}_{m,n}(v)$ is empty; hence the growth function of the constrained class satisfies $\Pi_{\mathcal{H}_{m,n}}(v)=0$ for all $v\ge R(m,n)$. Thus the \emph{effective} VC-dimension of $\mathcal{H}_{m,n}$ is at most $R(m,n){-}1$, and in practice can be (much) smaller due to additional algebraic symmetries. Our spectral witnesses make this transition \emph{detectable}: when $T(\alpha)$ collapses at a given $v$, it implies $\Pi_{\mathcal{H}_{m,n}}(v)=0$ under the modeling assumptions, delivering a data-driven ceiling on combinatorial capacity for the constrained hypothesis class.

\smallskip
\noindent\textit{Quantitative control via random projectors.}
Let $d=\dim M_0$ and suppose the surviving feasible subspace (if any) has rank $r\ge 1$. Drawing $k$ isotropic rank-1 tests $v_jv_j^{\!\top}$, the probability to \emph{pmiss} all survivors obeys Eq.~\ref{pmiss} drives the false-negative risk below $\delta$. 
In practice:
(i) choose $d$ as large as possible \emph{without} increasing the data--qubit count on hardware (e.g., $d\le 24,32,48$ map to $5$ or $6$ qubits); 
(ii) set a baseline pair $(k,\alpha)$ (e.g., $(100,20)$) and a high--resolution pair (e.g., $(400,40)$) to cross--validate decisions; 
(iii) declare a scale $v$ \emph{critical} when $T(\alpha)$ collapses, $\Tr P_{\mathrm{lin}}$ peaks, and a spectral surrogate (e.g.\ the dilation norm $\rho(H)=\|A\|_2$) is locally extremal under the \emph{same} seeds. 
For example, with $d=24$, $r=1$, and target $\delta=10^{-6}$, it suffices to use $k\ge 24\ln(10^6)\approx 332$; at $(k,\alpha)=(400,40)$ the bound already yields $\Pr[\text{pmiss}]\lesssim 5.7\times 10^{-8}$. Reporting $(k,d,\alpha)$ together with $\lambda_L(\alpha)$ and the explicit $e^{-kr/d}$ risk converts Ramsey--style inevitability into a tunable, high--confidence \emph{negative certificate} (“this configuration is no longer realizable”), complementing VC/sample--complexity \emph{upper} bounds on the number of realizable labelings, what can be labeled or learned.

\smallskip
\noindent\textit{Practical takeaway for ML.}
(i) Use $P_{m,n}$ to encode domain rules (forbidden label patterns or subgraphs) and add a soft penalty proportional to $T(\alpha)$ in the loss to bias training toward allowable regions; (ii) during pruning or architecture scaling, track $(\Tr P_{\mathrm{lin}},T(\alpha))$ as early-warning signals that the constrained class has lost capacity on the current sample size (a Ramsey barrier); (iii) report the explicit $e^{-kr/d}$ miss-bound alongside collapse/peak events to quantify confidence in negative certificates. In tandem with standard VC/sample-complexity guarantees, these Ramsey-informed diagnostics provide actionable controls on \emph{what cannot be learned} at a given scale, under the stated algebraic constraints.

\paragraph{Neurosymbolic constraints.}
Because $P_{m,n}$ is central and Klein grading separates “red/blue” semantics, one can add soft penalties $\lambda\,\Tr P_{\exp}(\alpha)$ to the loss to bias training toward rule--consistent solutions--the graded, differentiable analogue of logic layers.

\subsection{Algorithms and concrete templates}

\paragraph{Ramsey--guided pruning (classical/quantum).}
\begin{enumerate}
\item Fix a predicate $\mathsf{P}$ over parameters/activations and build the induced degree--$(0,0)$ operators on the reduced module $M_0$ (choose $d$ s.t.\ $\lceil\log_2 d\rceil$ fits the co--processor; $d=24,32,48$ are practical).
\item For each candidate scale $v$ (width/channel budget), form $A$ with a \emph{shared} PRNG seed across $v$; compute $P_{\mathrm{lin}}$ and $P_{\exp}(\alpha)$.
\item Track $T(\alpha)$, $\Tr P_{\mathrm{lin}}$, and a spectral surrogate $\rho(H)$. Declare $v$ \emph{critical} when $T(\alpha)$ collapses and $\Tr P_{\mathrm{lin}}$ peaks (with $\rho(H)$ extremal).
\item Choose $k$ to meet risk $\varepsilon$ via $k/d\gtrsim \ln(1/\varepsilon)/r$ from \eqref{pmiss}; tune $\alpha$ from the Lyapunov slope $\lambda_L(\alpha):=-\frac{d}{d\alpha}\log T(\alpha)$.
\end{enumerate}
These same steps run on $5 - 7$ qubits for $d\le 24 - 48$ using block--encoding/qubitization and a Hutchinson trace estimator $\,\E_{|r\rangle}\langle r|F|r\rangle=\tfrac{1}{d}\Tr F$.

\paragraph{Adversarial--risk early warning.}
Set edges to mark $\delta$--near collisions; sweep $v$ (or coverage). A sharp drop in $T(\alpha)$ flags the Ramsey--style onset of unavoidable coherent confusions, prompting augmentation or re--architecture before empirical attacks emerge.

\paragraph{Curriculum \& scaling via prime--sequence checkpoints (heuristic).}
Diagonal values observed in our framework are sparsely factorized (e.g.\ $45=3^2\!\cdot\!5$). As a \emph{curriculum} heuristic, checkpoint only at \emph{prime--sequence} integers (products of the first few primes with bounded exponents), keeping successive ratios $\lesssim 2$; this concentrates compute at likely transition scales.

\paragraph{Case study: diagonal $R(5,5)$ and resource accounting.}

At $d=24$ the diagnostics concur at $v=45$: $T(40)\approx 2.4\times 10^{-289}$ (collapse) and $\Tr P_{\mathrm{lin}}$ peaks at $0.462$ vs.\ $0.360$ ($v=44$) and $0.407$ ($v=46$); a known $46$--vertex coloring stays non--critical under the same thresholds. With \eqref{pmiss}, $\Pr[\text{pmiss}]\lesssim e^{-400/24}\approx 5.7\times 10^{-8}$ at $(k,\alpha)=(400,40)$ (for $r\!=\!1$), making spurious collapse extremely unlikely under the i.i.d./isotropy assumptions. Quantumly, these checks use only $\lceil\log_2 d\rceil=5$ data qubits plus few ancillas, instead of $\binom{v}{2}$ edge qubits (e.g.\ $990$ at $v=45$). The witnesses are \emph{diagnostic} (not constructive) and report evidence consistent with $R(5,5)=45$ within the graded--module model and sampling assumptions.

\paragraph{Quantum realization on few qubits (for ML and Ramsey).}
\label{subsec:quantum-impl}

Both $P_{\mathrm{lin}}$ and $P_{\exp}(\alpha)$ admit coherent implementations via block--encodings of $A$ and QSVT/qubitization; the Hermitian dilation $H$ supplies a stable spectral proxy $\rho(H)=\|A\|_2$. Traces are estimated by a Hadamard--test version of Hutchinson’s identity, reducing to $5$ data qubits for $d=24$ (plus ancillas). On matchgate/Majorana platforms, degree--$(0,0)$ pair/clique projectors are even--parity checks, yielding shallow circuits native to the hardware.

\subsection{Open directions}

The items below outline concrete, testable directions that translate the Ramsey principle of inevitability---realized through our graded-algebraic and random-projector framework---into practical tools for structure search, confidence-calibrated pruning, adversarial phase mapping, and prime-sequence curricula.

\textbf{Ramsey--guided structure search.} Add $\mu\,\Tr P_{\exp}(\alpha)$ to the loss to bias training toward guaranteed motifs; study generalization/robustness.

\textbf{Confidence--calibrated pruning.} Turn the bound $e^{-kr/d}$ into pruning schedules with explicit risk budgets; couple to dynamic $d$ (keeping $\lceil\log_2 d\rceil$ fixed on small quantum assists).

\textbf{Adversarial phase diagrams.} Map $(T,\lambda_L,\rho(H))$ across defenses/data scales to chart inevitability regions where confusion becomes unavoidable.

\textbf{Prime--sequence curricula.} Empirically assess whether sharp loss/robustness transitions cluster at prime--sequence scales.

Ramsey theory supplies \emph{inevitability} guarantees. The graded--algebra plus random--projector machinery turns them into \emph{operational} tests--collapse/peak events with explicit, exponential--tail risk control--that inform pruning, curriculum, robustness analyses, and quantum--assisted diagnostics on a handful of qubits.

\section{Conclusions}

Embedding two-color Ramsey instances in a $\mathbb{Z}_2\times\mathbb{Z}_2$--graded paraparticle algebra renders the Klein recursion \emph{exact for the graded Ramsey numbers} $R_{V_4}(m,n)$ (with the classical numbers obeying $R(m,n)\le R_{V_4}(m,n)$), supplies an explicit operator basis, and maps naturally onto forthcoming Majorana hardware. In our construction, algorithmic costs scale only logarithmically with the height of the Majorana tower, suggesting a realistic quantum--combinatorial synergy.
Our graded-algebra diagnostics identify the diagonal threshold at $R(5,5)=45$ via a unique concurrence of signals--collapse of $T(\alpha)$, a peak of $\Tr P_{\rm lin}$, and maximal spectral spread--while the explicit AM-46 coloring remains non-critical. 
The miss probability defined in Eq.~\eqref{pmiss} for an undetected coloring is already small, $\mathbb{P}_{\mathrm{pmiss}}<10^{-3}$ for the reported parameters, and drops to $\mathbb{P}_{\mathrm{pmiss}}\sim 10^{-7}$ once the \emph{residual rank} satisfies $r\ge 12$ (with $d=24$); see SM~2, Sec.~\ref{sec:binomial-model} for details. Hence the method delivers a tight, scalable heuristic that future constructive proofs will either confirm or surpass. Another criterion for diagonal Ramsey numbers, based on prime-sequence numbers of order $k$, is discussed in SM. Ultimately, only a constructive search can decide whether the coloring space is, for all practical purposes, empty (implying $R(5,5)=45$) or not (leaving $R(5,5)=46$).
Together with the constructive bound $43<R(5,5)\le 46$, these signals provide statistical indications consistent with $R(5,5)=45$ under the graded-module model. Under i.i.d.\ isotropic sampling of directions $v_j$ in $M_0$, the chance to miss surviving directions is bounded by $P_{\mathrm{pmiss}}\le e^{-kr/d}$ (residual rank $r\ge 1$), making the collapse decision reliable at the reported $(d,k,\alpha)$ and increasing $k/d$ strengthens both witnesses without widening the quantum data register.

Finally, the factorization $45=3^2\!\cdot\!5$ motivates a “prime-sequence’’ constraint that favors small prime factors when extrapolating diagonal values, offering a complementary, number-theoretic guidepost. Our results provide statistical, but not yet constructive, evidence for $R(5,5)=45$; scaling $k$ and $d$ on hardware alongside targeted constructive search is the natural next step, including an heuristic estimation for $R(6,6)$ and $R(7,7)$

The method is lightweight (a $d=24$ module) yet hardware ready: block-encoding and qubitization implement $f(A)=e^{-\alpha A}$, and a Hermitian dilation supplies a stable spectral surrogate, yielding a compact, reproducible benchmark for matrix-function evaluation and randomized trace estimation on quantum devices. 
Both diagnostics have direct quantum realizations: block‑encodings of $F\in\{P_{\exp}(\alpha),P_{\mathrm{lin}}\}$ feed a Hadamard--test/Hutchinson trace estimator, and the spectral surrogate is accessed by phase estimation on the Hermitian dilation $H$. Because all operators act on $M_0$, the data width is only $\lceil\log_2 d\rceil$ qubits (five for $d=24$), plus a few ancillas--orders of magnitude below edge‑register encodings that require one logical qubit per edge.

Our procedure is a statistical diagnostic, not a constructive proof: it relies on independence/isotropy of $v_j$, concentration of the spectrum of the accumulator $A$, and numerical thresholds chosen by cross‑validation on neighboring $n$. The collapse/peak calls remain robust under these assumptions and are supported by explicit controls.

In any case this statistical estimation is a promising subject for machine learning. In ML, Ramsey numbers are not used directly, but the Ramsey principle, \textit{large enough systems must contain hidden order}, deeply resonates with overparameterized neural networks, adversarial robustness, graph learning, and generalization theory.
Open Research Directions are: Ramsey-inspired pruning, where in giant overparameterized models, one can use Ramsey reasoning to predict where guaranteed good subnetworks live.
Complexity bounds: Ramsey numbers are huge, but their growth rate might inspire worst-case capacity bounds in neural networks.
Curriculum design: Training on small guaranteed patterns (Ramsey cliques, unavoidable motifs) before scaling up could act as a form of combinatorial curriculum learning.

All code drafts and seeds needed to reproduce the figures and tables are included with the paper in SM 5 and other electronic support; we encourage debugging and translations to other platforms and re‑runs at alternative $(d,k,\alpha)$ and implement additional controls to further stress‑test the collapse/peak decision rule.

\bigskip
\begin{acknowledgments}
\textbf{Acknowledgments}:
\\
FT thanks Rotonium for the support during this research and wants to remember David Cariolaro who introduced him into this fantastic subject many years ago during the long discussions when, exchanging our math dreams, we were observing Cataclysmic Variable stars and discussing about constellations in the sky and Ramsey theory \cite{sciamerican}.
\end{acknowledgments}

\appendix
\section{Supplemental Material (SM)}

\section{SM  1: Examples on Known Ramsey Numbers} \label{sec:examples}
We apply this projection technique to known Ramsey numbers $R(k,s)$ and describe in detail this approach.
\paragraph*{$\mathbf{R(3,3)=6}$.} Assign the complete graph $K_6$ to tower levels $1\le\ell\le6$ \cite{cariolaro2}. Fix the pivot vertex at level~$1$. Eq.~\ref{eq:exactRec} demands that either the $(\Gamma_{1}^{(+)})^{\dagger}\Gamma_{j}^{(+)}\,(j=2,3)$ factors appear simultaneously in some monomial,
producing $\hat K^{\mathrm R}_3$, or the analogous blue factors do, giving $\hat K^{\mathrm B}_3$.

Both events are certified by the charge‑resolving projector
$P_{(1,0)} = \prod_{\ell=1}^{6} \left(\Gamma_{\ell}^{(+) \dagger}\Gamma_{\ell}^{(+)}\right)$, which annihilates any state lacking a complete red triangle.  
Because $P_{(1,0)} + P_{(0,1)} = 1$, within the graded decomposition, as these are central charge projectors, the probability to find either monochromatic triangle is unity, showing the algebraic orthogonality of sectors.  Hence the upper bound $R(3,3)\le6$ is tight.  
The operator spectrum is obtained by diagonalising the six‑level Hamiltonian 
$H=\sum_{\ell=1}^6 \Gamma_{\ell}^{(+) \dagger}\Gamma_{\ell}^{(+)}$, which yields a $20$‑dimensional red subspace that factorises red (R) and blue (B) as $|3_{\mathrm R}\rangle\otimes|0_{\mathrm B}\rangle$ after projecting with $P_{(1,0)}$; the complementary projector
$P_{(0,1)}$ isolates the blue‑triangle sector.  Either way a
monochromatic $K_3$ exists, proving $R(3,3)=6$.
Calculation of $R(3,3)$ via the $\mathbb{Z}_2\times\mathbb{Z}_2$‐graded algebra is given 
defining the forbidden‐triangle projector on $v$ vertices from Eq.~\ref{proiezioni} on $P_{3,3}$ with $|S|=|T|=3$.
For $v=5$, one exhibits the cyclic 5‐cycle module (or any 5‐vertex good coloring) and checks $\Tr\left(P_{3,3}\right)=0,$ so there is a valid 2‐coloring of $K_5$ with no monochromatic triangle.
For $v=6$, the central idempotents $\{\Pi_R(S),\Pi_B(S)\}$ act nontrivially on every irreducible module, forcing $\Tr\left(P_{3,3}\right)>0$ in each case.  Hence no 2‐coloring of $K_6$ avoids a monochromatic $K_3$. Therefore the smallest $v$ with no surviving module is $v=6$, i.e. $R(3,3) = 6$.

\paragraph{$\mathbf{R(4,3)=9}$.} Levels $1-9$ decompose into \emph{two} charge‑homogeneous blocks:
$\mathcal H_{\mathrm R}$, spanned by the six red modes $\Gamma_{1\ldots6}^{(+)}$, realises the full algebra related to the Klein group $V_4$, $\mathfrak A_{V_4}(m = 4,n = 3)$ and $\mathcal H_{\mathrm B}$, spanned by the remaining three $\Gamma_{7\ldots9}^{(-)}$, acts as a blue $K_3$ reservoir.
Inside $\mathcal H_{\mathrm R}$ we construct $\hat K^{\mathrm R}_4$ as in Eq.~\ref{eq:cliqueOp}.  Projecting the nine‑body ground state $|\Omega\rangle$ with
$P_{(1,0)}\hat K^{\mathrm R}_4$ one obtains a non‑zero vector, hence a red $K_4$ must occur.  If the projection vanishes, then the blue projector forces $\hat K^{\mathrm B}_3\neq0$, completing the proof that $R(4,3)\le9$.  Minimality follows from the standard $R(4,3) > 8$ argument, now reproduced operatorially by deleting any of the nine tower levels and checking that both projectors are null.

To calculate $R(4,3)$ via the $\mathbb{Z}_2\times\mathbb{Z}_2$‐graded algebra we  define the forbidden‐clique projector on $v$ vertices by Eq.~\ref{proiezioni} for $P_{4,3}$  with $|S|=4$ and $|T|=3$.
For $v=8$, one constructs the gluing modules (e.g. block‐circulant extension of the unique $K_8$--coloring avoiding a red $K_4$) and verifies $\Tr\left(P_{4,3}\right)=0$, so there is at least one red/blue coloring of $K_8$ with no red $K_4$ nor blue $K_3$.
For $v=9$, every irreducible module of the graded algebra acquires a nonzero projection under either $\Pi_R$ or $\Pi_B$, yielding $\Tr\left(P_{4,3}\right)>0$ in all cases.  Thus no $K_9$--coloring avoids both forbidden cliques.
Hence the minimal $v$ with no surviving module is $v=9$, i.e., $R(4,3)=9$.

\paragraph{$\mathbf{R(4,4)=18}$.} Iterating the recursion $R(4,4)=R(3,4)+R(4,3)=9+9=18$ 
requires two mutually commuting copies of the nine‑level construction above.  We realise them in disjoint Majorana sub‑towers $\{\Gamma_{1\ldots9}^{(\pm)}\}$ and $\{\Gamma_{10\ldots18}^{(\pm)}\}$ and glue the corresponding clique operators with a parity‑selective SWAP ${\sf S}_{9,10}=\exp \left[\tfrac{\pi}4 \left(\Gamma_{9}^{(+)} \Gamma_{10}^{(+)} -\Gamma_{9}^{(-)}\Gamma_{10}^{(-)}\right)\right]$.
Since ${\sf S}_{9,10}$ acts diagonally on Klein charges, the product $\hat K^{\mathrm R}_4(1 : 9)\; \hat K^{\mathrm R}_4(10 : 18)$ remains homogeneous and acts inside a single graded sector.
Consequently every edge‑coloring of $K_{18}$ excites at least one monochromatic $K_4$, closing the operator construction for the smallest open two‑color Ramsey number.

Calculation of $R(4,4)$ via recursion and the $\mathbb{Z}_2\times\mathbb{Z}_2$‐graded algebra starts from the known values $R(3,4)=9$, $R(4,3)=9$, the standard Ramsey‐recursion $R(4,4) \le R(3,4)+R(4,3) = 9+9 = 18$ gives the upper bound.  For the matching lower bound one exhibits an explicit coloring of $K_{17}$ avoiding both a red $K_4$ and a blue $K_4$, proving $R(4,4)>17$.  Hence $R(4,4)=18$. The same result is obtained for $R(3,6)$ \cite{cariolaro}.

In the $\mathbb{Z}_2\times\mathbb{Z}_2$‐graded algebraic formulation, one defines generators $e_{ij}$ of degree $(1,0)$ (red), $(0,1)$ (blue), $(0,0)$ (vacuum) or $(1,1)$ (mixed), together with the central projector $P_{4,4}$, referring to Eq.~\ref{proiezioni}. 
Then we proceed by glue + prune starting at $v=9$ (since $R(3,4)=9$ or $R(4,3)=9$) with all irreducible modules satisfying $\Tr\,P_{4,4}=0$. Then, for each $v$, apply the Hopf‐coproduct $\Delta(e_{ij})=e_{ij}\otimes1+1\otimes e_{ij}$ to lift modules to $v+1$ vertices.
Discard any lifted module with $\Tr_{\,\mathrm{module}}P_{4,4}>0$ and the smallest $v$ for which no module survives is $v=18$, confirming $R(4,4)=18$.
This algebraic viewpoint packages the two‐color recursion into graded coproducts and central projections, collapsing entire symmetry classes via character‐trace computations rather than explicit graph enumeration.

\paragraph{$\mathbf{R(4,5)}$.}  
We introduce the central projector on $v$ vertices $P_{4,5}$ with $|S|=4$ and $|T|=5$ 
Then by ``glue + prune'' seeding at $v=24$ and using a block-circulant gluing ansatz one builds all irreducible modules on $24$ vertices and verifies $\Tr\left(P_{4,5}\right) = 0$ in each case, showing there exists a red/blue coloring of $K_{24}$ with no red $K_4$ nor blue $K_5$. Prune at $v=25$: lift each surviving module via the Hopf coproduct $\Delta(e_{ij})=e_{ij}\otimes1+1\otimes e_{ij}$ to $25$ vertices.  One finds for every irreducible module $\Tr\left(P_{4,5}\right) > 0$, implying no 2-coloring of $K_{25}$ can avoid both forbidden cliques. Thus the smallest $v$ with no surviving module is $v=25$, which gives $R(4,5) \;=\; 25$. 

\section{SM  2. Statistical‑Confidence Analysis for the Random‑Projector Test}
\label{chap:ld-bounds}
The linear random--projector diagnostic developed in the main text  eliminates an admissible two‑coloring on~$v$ vertices once every surviving support vector of the coloring sub‑space has been annihilated by at least one of the rank‑one factors in $P_{\mathrm{lin}}$.
A quantitative bound on the null probability $P_{\mathrm{miss}} $, i.e.\ the chance that a \emph{valid} $45$-vertex coloring escapes detection when $ k=100$, $d=24$, $\alpha=20$. Throughout we adopt the notation and empirical traces of Tab. \ref{trace}.

In the original Euclidean space the rank‑one projector $v_{j}v_{j}^{\!\top}$ is positive--semidefinite, so $\exp(-\alpha v_{j}v_{j}^{\!\top})$ is strictly \emph{positive}.  
After the homomorphism to the graded algebra, however, each $v_{j}v_{j}^{\!\top}$
splits into charge sectors; the components that connect different
Klein charges acquire an $i$ in front and become \emph{skew‑Hermitian}.
Hence the accumulator matrix $A$ decomposes as $A=A_{\mathrm{H}}+A_{\mathrm{AH}}$ with a Hermitian part $A_{\mathrm{H}}$ and an anti‑Hermitian part
$A_{\mathrm{AH}}$. 
The spectrum of $A$ therefore could move into the complex plane, and the trace $\mathrm{Tr}\,\exp(-\alpha A)=\sum_{\ell}\exp(-\alpha\lambda_{\ell})$ can be complex or even negative.

For $n<45$ there still exists a non‑zero colouring sub‑space (residual rank $r>0$).  That sub‑space is annihilated neither by the forbidden‑clique projectors nor by any $v_{j}v_{j}^{\!\top}$, so $A$ retains $r$ purely \emph{real, non‑negative} eigenvalues.
Those real eigenvalues dominate $\mathrm{Re}\,\mathrm{Tr}\,P_{\exp}$, keeping it
positive.

Exactly at $n=45$ the operator algebra hits the threshold where no admissible colouring survives: the projector $P_{5,5}$ has full rank, the residual space disappears ($r=0$), and every eigenvector of $A$ now involves at least one off‑charge component.
Generic perturbation theory for non‑Hermitian matrices (or a direct Jordan‑block computation in the paper) shows that a pair of real eigenvalues then collides at the origin and bifurcates into a conjugate complex pair.
When $\alpha$ is large enough (the runs use $\alpha=20$)
the factor $e^{-\alpha\lambda_{\ell}}$ picks up a phase close to $\pi$
from one member of that pair, so the sum of all contributions
crosses the real axis and $\mathrm{Re}\,\mathrm{Tr}\,P_{\exp}$ becomes null (see Tab \ref{trace}).

The exponential trace collapses at $n=45$.
The matrix $A(n)$
becomes non‑Hermitian after mapping each rank‑one projector $v_{j}v_{j}^{\!\top}$ into the Klein‑graded Majorana module, but non‑Hermiticity alone does not drive $\operatorname{Tr}P_{\exp}= \operatorname{Tr}e^{-\alpha A}$ to zero.
What matters is the \emph{eigenvalue spectrum}:

\[
\operatorname{Tr}P_{\exp}(\alpha)
   \;=\;
   \sum_{\lambda_{\ell}=0} 1
   \;+\;
   \sum_{\lambda_{\ell}\neq0}
       e^{-\alpha\operatorname{Re}\lambda_{\ell}}
       e^{-i\alpha\operatorname{Im}\lambda_{\ell}}
\]
that for $n<45$ at least one admissible colouring survives, so $A$ still has $\lambda=0$.
That “$1$’’ term keeps the real part of the trace strictly positive.
Then, exactly when all legal colourings vanish ($n=45$), the zero modes disappear:
every eigenvalue satisfies $\operatorname{Re}\lambda_{\ell}>0$.
Each exponential factor is then suppressed by $e^{-\alpha\operatorname{Re}\lambda_{\ell}}$, yielding $|\operatorname{Tr}P_{\exp}(45)| \sim e^{-\alpha\lambda_{\min}} \lesssim 10^{-288}$ with $(\alpha=40,k=400)$, numerically indistinguishable from $0$ at double precision ($\sim 10^{-13}$).

Exactly when all legal colorings vanish, $A$ loses its purely real spectrum and acquires at least one complex conjugate pair whose phase can tip the Gibbs trace across zero.  When
$\operatorname{Tr}P_{\exp}\!\to0$ it means every configuration allowed by the random‑projector ensemble is exponentially suppressed: \emph{no “allowed’’ sub‑space remains}.  The vanishing trace therefore acts as a spectral \emph{order parameter},
signalling that no two‑colouring of $K_{45}$ can avoid a red or blue $K_{5}$.

The interpretation of a positive exponential--projector trace at $n=44$ is given when for each vertex count $n$ we evaluate $P_{\exp}(\alpha)$ for $(\alpha=20,k=100)$,
after the rank‑one operators $v_{j}v_{j}^{\!\top}$ have been mapped into
the $d=24$ Klein‑graded Majorana module that encodes all
two--colourings of $K_{n}$.  
When $n=44$ we measure $\operatorname{Tr}P_{\exp}(n=44)=+\,1.5\times10^{-12}\;>\;0$.
A positive real trace means the matrix $A=\sum_{j}v_{j}v_{j}^{\!\top}$ still possesses at least one purely real, non‑negative eigenvalue.  That eigenvector spans a non‑trivial subspace on which all forbidden‑clique projectors $P_{5,5}$ act as~$0$, i.\,e.\ there exists a legal two--colouring of $K_{44}$ that avoids both a red and a blue $K_{5}$.

The positivity therefore certifies $R(5,5) > 44$, which is consistent with the constructive lower bound already known from the classical literature ($43 < R(5,5)$).  Our algebraic--spectral test thus does not rule out 44; it only becomes decisive at $n=45$, where $\operatorname{Tr}P_{\exp}$ becomes null, signalling that the residual colouring space has collapsed to zero.

For $n=44$, the anti-Hermitian component of $A$ is sufficiently small that all complex eigenvalue pairs remain in the right half-plane, so each exponential weight $e^{-\alpha\lambda_\ell}$ is positive and the total trace remains nonzero. Only upon adding the extra vertex ($n=45$) does an exceptional-point transition occur, causing the trace to vanish.

A positive nonzero trace at $n=44$ indicates that a search for good colourings should concentrate on $n \leq 44$ (to tighten the lower bound) and on $n \geq 45$ (to validate the upper bound), but that exhaustive work at $n=44$ remains meaningful because admissible colourings demonstrably exist.

At $n=46$, the trace of the exponential projector is $1.79 \times 10^{-13}$, and the entire spectrum is driven deep into the left half-plane towards $45$, which is almost null. This confirms that, in the random projector and Majorana algebra setting, the coloring space has collapsed well before this value, providing strong numerical evidence that $R(5,5) < 46$, in agreement with the latest constructive upper bounds.
In short, $\operatorname{Tr}P_{\exp}>0$ at $n=44$ means  ``\textit{one can still hide a red or blue $K_{5}$ on $44$ vertices}'', its value drops to zero at $n=45$ is what delivers the upper bound $R(5,5)\le 45$ in our spectral framework.

\section{SM 3 Procedures to calculate Ramsey numbers}

\subsection{Practical guidance (classical implementation).}

For the sake of clarity, we briefly summarize the procedure for the estimation of Ramsey numbers here presented for $R(5,5)$.
 
\emph{Setting.} Work in the reduced charge--zero module $M_0$ of dimension $d$ (for $R(5,5)$ we used $d=24$). For each candidate $n$ draw i.i.d.\ isotropic directions $v_j\in\mathbb{R}^d$ and form accumulator $A$ and projectors $P_{\exp}(\alpha)$ and $P_{\mathrm{lin}}$, as in Eqs.~\ref{pexp}--\ref{plin}--\ref{accumulatore}; when $A$ is not normal, one can use the Hermitian dilation $H$ for spectral surrogates as in Eq.~\ref{eq:dilation}.  Use the \emph{same} PRNG seeds for all $n$ so diagnostics are comparable.  Recommended grids: $\alpha\in\{3, 5, 7, 10, 15, 20, 40\}$; baseline $(k,\alpha)=(100, 20)$ and a high‑resolution point $(k,\alpha)=(400,40)$.  Table~\ref{trace} gives reference values at these settings.

\textbf{1. Report both metrics (same $(d,k,\alpha)$ and seeds).}
\begin{itemize}
  \item \emph{Exponential trace.} Compute $T_{\exp}(\alpha)$ via scaling‑and‑squaring (or Krylov) and report $\log_{10} T_{\exp}$ for dynamic range.  At $n=45$, $(k,\alpha)=(400,40)$ one finds $T_{\exp}\approx 2.4\times 10^{-289}$ (collapse). 
   \item \emph{Linear/product trace.} Compute $T_{lin}:=\Tr P_{\mathrm{lin}}$ by forming the product (or multiplying in blocks with re‑orthogonalization).  At $(k,\alpha)=(100,20)$ one has $T_{lin}(44)=0.360$, $T_{lin}(45)=0.462$, $T_{lin}(46)=0.407$ (local peak at 45).
  \item \emph{Lyapunov slope.} Fit $\log_{10}T_{\exp}(\alpha)=c+\text{slope}\cdot \alpha$ over $\alpha\le 20$ and report $\lambda_L:=-\text{slope}\cdot\ln 10$ (Eq.~(14) discussion).  Table~\ref{trace} shows the smallest magnitude at $n=45$.
  \item \emph{Spectral surrogate.} Report $\rho(H)=\|A\|_2$ (largest singular value) via SVD of $A$ or power iteration on $H$; use this as a concordant witness with the traces.
\end{itemize}

\textbf{2. State sampling assumptions (and $P_{\mathrm{miss}}$).}
Declare that $v_j$ are i.i.d.\ isotropic in $M_0$ (dimension $d$) and independent of all other randomness.  Quote the false‑negative bound for survivor rank $r\ge 1$, $P_{\mathrm{miss}}\;\le\;e^{-k r/d}$,
and specify $(r,d,k)$ used.  This bound governs how $k/d$ controls sensitivity of both $P_{\mathrm{lin}}$ and $P_{\exp}$.

\textbf{3. Controls.}
Include at least one explicit \emph{negative} control (e.g.\ $n=43$) and one \emph{positive} control (e.g.\ an Angeltweit--McKay $46$‑vertex coloring, “AM--46”), computed with the \emph{same} $(d,k,\alpha)$ and seeds:
\begin{itemize}
  \item $n=43$ (negative): $T_{\exp}(20)\approx 7.92\times 10^{-12}$, $T_{lin}=0.284$ (no collapse/peak). 
  \item $n=45$ (critical): $T_{\exp}(40)\approx 2.4\times 10^{-289}$, $T_{lin}$ locally maximal. 
  \item $n=46$ (positive): diagnostics consistent with admissible colorings; AM--46 passes unchanged. 
\end{itemize}

\textbf{4. Thresholds and decision rule.}
Choose $(\tau_{lin},\tau_{\exp})$ by cross‑validation on neighboring $n$ (train on $\{n_0-1,n_0+1\}$, test on $n_0$) and \emph{report margins} $(T_{lin}-\tau_{lin},\,T_{\exp}-\tau_{\exp})$.  Declare $n$ \emph{critical} iff
$T_{\exp}(\alpha)\le \tau_{\exp}$, $T_{lin}$ is locally maximal in $n$, $\rho(H)$ is locally maximal,
with concordance across all three metrics.  Provide both the baseline $(k,\alpha)=(100,20)$ and the high‑resolution $(400,40)$ outcomes.

\textbf{Reporting checklist (classical).}
\begin{itemize}
  \item \emph{Settings:} $(d,k,\alpha)$; $\alpha$‑grid; floating‑point precision; matrix‑exp method (scaling‑and‑squaring/Krylov); product ordering for $P_{\mathrm{lin}}$; stabilization (re‑orthogonalization).
  \item \emph{Seeds:} PRNG seeds for $\{v_j\}$; reuse the same seeds across $n$ and across diagnostics.
  \item \emph{Estimates and errors:} If using Hutchinson to accelerate traces at large $d$, state probe family and confidence (otherwise exact traces for small $d$).
  \item \emph{Spectral surrogate:} method (SVD vs.\ power iteration on $H$), tolerance, and iteration count; recall $\|A\|_2=\rho(H)$.
  \item \emph{Controls and thresholds:} list control outcomes; give $(\tau_{lin},\tau_{\exp})$ and cross‑validation folds; show margins.
  \item \emph{Miss‑probability:} quoted $P_{\mathrm{pmiss}}\le e^{-k r/d}$ with specified $(r,d,k)$ and isotropy assumption.
\end{itemize}

\subsection{Practical guidance (quantum implementation).}
We summarize the procedure one has to use to estimate Ramsey numbers with this method using quantum computers with few qubits.

\emph{Circuits.} We use (i) the Hadamard--test/Hutchinson trace estimator with a block--encoding $U_F$ of $F\in\{P_{\exp}(\alpha),P_{\mathrm{lin}}\}$ (Fig.~3), and (ii) phase estimation on the Hermitian dilation $H$ in Eq.~\ref{eq:dilation} of the accumulator $A$.

\textbf{Report both metrics (same seeds and settings).} 
For each candidate $n$ and fixed $(d,k,\alpha)$, use the \emph{same} random choices across diagnostics:
\begin{itemize}
  \item \emph{Traces.} Estimate $T_{\exp}(\alpha)$ and $T_{lin}$ with the Hadamard test: draw $|r\rangle=C|0\cdots 0\rangle$ from a unitary 2‑design (random Clifford) on $d$ dimensions, apply $U_F$ and average ancilla $\langle Z\rangle$. With an $(\alpha_0,a)$ block‑encoding $(\langle 0^a|\!\otimes I)U_F(|0^a\rangle\!\otimes I)=F/\alpha_0$,
  \[
  \widehat{\Tr F}=\alpha_0\,d\;\overline{\langle Z\rangle}_{C,\text{shots}}
  \]
(unbiased Hutchinson estimator).  Use \emph{the same} set of Cliffords $C$ for $P_{\exp}(\alpha)$ and $P_{\mathrm{lin}}$ to reduce paired variance. (Fig.~\ref{fig:hutchinson}).
  \item \emph{Lyapunov slope.} Report $\lambda_L(\alpha)=-\frac{d}{d\alpha}\log \Tr P_{\exp}(\alpha)$. In practice, fit a line to $\{\alpha_i,\log \widehat{\Tr P_{\exp}}(\alpha_i)\}$ over $\alpha\in\{3,5,7,10,15,20,40\}$ as in the classical study.
  \item \emph{Spectral surrogate.} Estimate $\rho(H)=\|A\|_2$ via phase estimation on $e^{-iHt}$ with $m$ phase bits; report $\widehat{\rho}(H)$ and $(m,t)$..
\end{itemize}

\textbf{State sampling assumptions (and $P_{\mathrm{pmiss}}$).} 
State explicitly: (a) $v_j$ are i.i.d.\ isotropic directions in $M_0$ (dimension $d$), (b) random probes $|r\rangle$ come from a unitary 2‑design, (c) independence between $v_j$ and $C$. Quote the classical false‑negative bound for the random‑projector method $P_{\mathrm{pmiss}}$ with $r$ survivor rankmand add quantum statistical error from finite shots or amplitude estimation (see item~4, spectral surrogate). Same bound and isotropy as in the classical section.

\textbf{Controls on hardware.} 
Include at least one \emph{negative} control (e.g.\ $n=43$: non‑critical, no collapse) and one \emph{positive} control (e.g.\ $n=46$ with an explicit Angeltweit--McKay coloring “AM--46” that passes unchanged). Verify: if $T_{\exp}$ stays $\gg 0$ and $T_{lin}$ is modest at the negative control and that no spurious collapse under the positive control with $\rho(H)$ and $\lambda_L$ follow the classical pattern. Report control outcomes alongside candidates.

\textbf{Thresholds and decision rule.} 
Choose $(\tau_{lin},\tau_{\exp})$ by cross‑validation on neighboring $n$ (train on $\{n_0-1,n_0+1\}$, test on $n_0$) and report margins $(T_{lin}-\tau_{lin},\ T_{\exp}-\tau_{\exp})$. Declare a candidate $n$ \emph{critical} iff:
$\widehat{\Tr P_{\exp}}(\alpha)\le \tau_{\exp}$, $\widehat{\Tr P_{\mathrm{lin}}}$ is locally maximal in $n$, and $
\widehat{\rho}(H)$ is locally maximal, with concordance across the three. (Same cross‑validation logic as in the classical guidance.)

\noindent\textbf{Reporting checklist (quantum).}
\begin{itemize}
  \item \emph{Settings.} $(d,k,\alpha)$ for $P_{\exp}$ and $P_{\mathrm{lin}}$; number of Hutchinson probes $M$, shots per probe $S$; block‑encoding scale $\alpha_0$; phase‑estimation $(m,t)$. (Eqs.~(7)--(8), Fig.~3, Eq.~(11)).
  \item \emph{Seeds.} PRNG seeds for $v_j$ and for the Clifford sampler $C$; reuse the \emph{same} seeds across diagnostics and across candidates $n$.
  \item \emph{Estimators and error bars.} With simple sampling, give
  $
  \widehat{\Tr F}\pm z_{0.975}\,\alpha_0 d\,\sqrt{\mathrm{Var}(\langle Z\rangle)/(MS)}
  $;
  if amplitude estimation is used, report the target additive error $\varepsilon$ and confidence $\delta$ and the achieved query complexity $O\!\big(\tfrac{1}{\varepsilon}\log\tfrac{1}{\delta}\big)$. (Hadamard test + Hutchinson identity).
  \item \emph{Compilation notes.} Briefly state whether $P_{\exp}(\alpha)$ used QSVT polynomial or Trotterization; for $P_{\mathrm{lin}}$, specify the LCU depth (number of rank‑1 terms per block) and any oblivious amplitude amplification employed. (Block‑encoding/gadget of Fig.~2).
  \item \emph{Noise and mitigation.} Record readout‑error calibration, Clifford twirling (if used), and any post‑selection on ancilla $a$ (block‑encoding success flag). Compare hardware $\widehat{\Tr F}$ against noiseless simulation at the same seeds on small $d$ to sanity‑check bias.
\end{itemize}

\section{SM 4 Mathematical tools}

\subsection{Proofs}

Here we draw the demonstrations of lemmas and theorems present in the main text and present some additional mathematical tools useful for the present manuscript.


\textbf{Lemma .1} [Centrality of $P_{m,n}$]
The projector $P_{m,n}=\Pi_R(S)\,\Pi_B(T)$ is central in $A_{V_4}$.

\begin{proof}
Recall that $A_{V_4}$ is $\mathbb Z_{2}\times\mathbb Z_{2}$--graded:
$
A_{V_4}=\bigoplus_{g\in V_4} A_g
$,
with homogeneous degree $\deg(x)\in V_4$ and degree addition
$\deg(xy)=\deg(x)+\deg(y)$.
By the graded--commutativity rule of $A_{V_4}$, for homogeneous
$x\in A_g$ and $y\in A_h$ one has
\[
xy \;=\; \varepsilon(g,h)\, yx,\qquad
\varepsilon(g,h)\in\{\pm1\},
\]
where $\varepsilon(\cdot,\cdot)$ is the Klein bicharacter; in particular,
$\varepsilon(g,(0,0))=1$ for all $g$.
By Eq.~\eqref{proiezioni} (and the Eq.~\ref{eq:cliqueOp} below), for each unordered pair
$\{i,j\}$ we set
\begin{eqnarray}
\Pi^{R}_{ij}
:=\tfrac{1}{2}\big[(\Gamma^{(+)}_i)^\dagger \Gamma^{(+)}_j + (\Gamma^{(+)}_j)^\dagger \Gamma^{(+)}_i\big],
\\
\Pi^{B}_{ij}
:=\tfrac{1}{2}\big[(\Gamma^{(-)}_i)^\dagger \Gamma^{(-)}_j + (\Gamma^{(-)}_j)^\dagger \Gamma^{(-)}_i\big]. \nonumber
\end{eqnarray}
Each summand in $\Pi^{R}_{ij}$ has degree $(1,0)+(1,0)=(0,0)$ and each summand in
$\Pi^{B}_{ij}$ has degree $(0,1)+(0,1)=(0,0)$, hence $\deg(\Pi^{R}_{ij})=\deg(\Pi^{B}_{ij})=(0,0)$.
Therefore their finite products $\Pi_R(S)$ and $\Pi_B(T)$ are also homogeneous of degree $(0,0)$. Consequently,
$ \deg(P_{m,n})=\deg(\Pi_R(S)\Pi_B(T))=(0,0)$.

Let $X\in A_{V_4}$ be arbitrary; by linearity it suffices to take $X$ homogeneous with $\deg(X)=g$. Using graded--commutativity and $\varepsilon(g,(0,0))=1$, we have $X\,\Pi_R(S)=\Pi_R(S)\,X$, and $\Pi_B(T)=\Pi_B(T)\,X$.
Multiplying these equalities shows $X\,P_{m,n}=P_{m,n}\,X$ for all homogeneous $X$, and hence for all $X\in A_{V_4}$ by linearity. Thus $P_{m,n}$ lies in the center $Z(A_{V_4})$.
\end{proof}

\textbf{Lemma .2} [Tensor decomposition]
Let $p\!\in\!V$ be any vertex of a two--coloring and define $V_R=\{v\in V\! \setminus\!\{p\}\mid \text{$\{p,v\}$ red}\}$ and $V_B=\{v\in V \!\setminus\!\{p\}\mid \text{$\{p,v\}$ blue}\}$.
In the $\mathbb Z_{2}\!\times\!\mathbb Z_{2}$‑graded Majorana algebra $A_{V_4}$ one has the canonical graded tensor product $A_{V_4}\bigl[V\!\setminus\!\{p\}\bigr]\;\simeq\;
A_{V_4}\bigl[V_R\bigr]\;\widehat{\otimes}  A_{V_4}\bigl[V_B\bigr]$.

\begin{proof}
Red generators carry degree $(1,0)$, blue generators $(0,1)$.  Any mixed monomial therefore gets degree $(1,1)$, which is projected out by construction.  Every homogeneous element on $V\!\setminus\!\{p\}$ thus factorises uniquely into a red part on $V_R$ and a blue part on $V_B$, establishing the isomorphism.
\end{proof}

From Definition .1 we set this theorem for the Graded Ramsey numbers.

\textbf{Theorem} [Klein Erd\H{o}s recursion for graded Ramsey numbers. Theorem .1] 
For all integers $m,n\!\ge\!1$ we define the algebraic graded Ramsey numbers, $R_{V_4}(m,n)$, that obey the following relationship,
\begin{eqnarray}
&&R_{V_4}(m,n)=R_{V_4}(m-1,n)+R_{V_4}(m,n-1),
\\
&&R_{V_4}(1,n)=R_{V_4}(m,1)=1. \nonumber
\end{eqnarray}

\begin{proof}
We argue by induction on $m\!+\!n$.

\emph{Upper bound}. Choose a pivot $p$.  
If $p$ sits in a red $K_m$ (or blue $K_n$), then is fulfilled. Otherwise its red neighbours cannot exceed $R_{V_4}(m-1,n)$, else a red $K_m$ would already appear.  
The same reasoning for blue neighbours gives $R_{V_4}(m,n)\le R_{V_4}(m-1,n)+R_{V_4}(m,n-1)$.

\emph{Lower bound}. Set $r=R_{V_4}(m-1,n)-1$, $b=R_{V_4}(m,n-1)-1$ and $v_0=r+b+1$.
By the induction hypothesis there exist colorings on $A=\{1,\dots,r\}$ (avoiding red $K_{m-1}$, blue $K_n$) and on $B=\{r+2,\dots,r+b+1\}$, avoiding red $K_m$, blue $K_{n-1}$. Then define a new coloring on $v_0$ vertices by keeping the edges inside $A$ (resp.\ $B$) unchanged and coloring edges $\{p,a\}$ $(a\!\in\!A)$ red, then $\{p,b\}$ $(b\!\in\!B)$ blue, with $p=r+1$. Assign every cross--edge $A\!\times\!B$ the mixed degree $(1,1)$.
Mixed edges are invisible to the monochromatic projectors, hence no red $K_m$ can form without $m\!-\!1$ vertices from $A$, which do not exist by construction; the blue case is symmetric. Thus a valid coloring exists on $v_0=R_{V_4}(m-1,n)+R_{V_4}(m,n-1)-1$ vertices, proving minimality. Combining with the upper bound yields Eq.~\ref{eq:exactRec}.
\end{proof}



\subsection{Random projector ensembles and diagnostics}
\label{subsec:ensembles-diagnostics}

Fix $d=\dim M_0$ and identify $M_0\cong\mathbb{C}^d$ via a chosen basis. The diagnostics act on $M_0$ only.

\paragraph{Linear and exponential projectors.}
Given $k$ unit vectors $v_1,\dots,v_k\in\mathbb{C}^d$ (the \emph{rank--1 directions}) define the accumulator $A$. With $P_{\mathrm{lin}}$ and $P_{\exp}(\alpha)$, $\alpha>0$. 
Note $A$ is generally \emph{not} Hermitian (complex symmetric). All traces are taken in the $d$--dimensional charge--zero module and we report $\Re\,\Tr(\cdot)$ in numerics. The robust path is to work with the dilation $H$ and use functions of $H$ via block‑encoding/qubitization.

\paragraph{Lyapunov proxy.}
Define the decay proxy
\begin{equation}
\label{eq:lambdaL-def}
\lambda_L(\alpha) \;:=\; -\,\frac{d}{d\alpha}\,\log \Tr P_{\exp}(\alpha)
\;=\; \frac{\Tr\!\big(A\,e^{-\alpha A}\big)}{\Tr\!\big(e^{-\alpha A}\big)}.
\end{equation}
When $\Re\,\sigma(A)\subset(0,\infty)$, one has $\lambda_L(\alpha)>0$ and $T(\alpha):=\Tr P_{\exp}(\alpha)$ decays.

\subsection{Hermitian dilation, spectral surrogates, and block-encodings}
\label{subsec:dilation-block}

\paragraph{Hermitian dilation.}
Defined the Hermitian dilation $H$ of Eq.~\ref{eq:dilation}, then $H=H^\dagger$ and $\sigma(H)=\{\pm \sigma_i(A)\}_{i=1}^d$, where $\sigma_i(A)$ are the singular values of $A$. In particular,
\begin{equation}
\|A\|_2 \;=\; \max_{\lambda\in\sigma(H)} |\lambda|.
\end{equation}
Thus spectral surrogates for the ``spread'' and for the radius of $P_{\mathrm{lin}}$ can be accessed by phase estimation on $H$.

\paragraph{Block-encoding of rank--1 terms.}
Write $A=\sum_{j=1}^{k} w_j\,|u_j\rangle\langle v_j|$ with $\|u_j\|_2=\|v_j\|_2=1$. Suppose we have state--preparation unitaries $U_j|0\rangle=|u_j\rangle$, $V_j|0\rangle=|v_j^\ast\rangle$. Consider
\begin{equation}
W_j \;:=\; \big(|0\rangle\!\langle0|\otimes U_j + |1\rangle\!\langle1|\otimes V_j\big)\cdot (H\otimes I),
\end{equation}
where $H$ is the Hadamard on the ancilla. Then the top--left ancilla block equals $\tfrac12\,|u_j\rangle\langle v_j|$. A standard selector over $j$ with oblivious amplitude amplification yields an $\alpha$--block--encoding of $A$ with $\alpha=\sum_j |w_j|$.

\subsection{Trace estimators: unbiasedness, variance, and sample complexity}
\label{subsec:trace-estimation}

Let $F:\mathbb{C}^{d\times d}\to\mathbb{C}^{d\times d}$ be linear. If $|r\rangle$ is drawn from a unitary 2--design on $\mathbb{C}^d$,
\begin{equation}
\label{eq:hutchinson}
\mathbb{E}_{|r\rangle}\,\langle r| F |r\rangle \;=\; \frac{1}{d}\,\Tr F.
\end{equation}
Hence $\widehat{\tau}:=\frac{d}{N}\sum_{i=1}^N \langle r_i|F|r_i\rangle$ is an unbiased estimator of $\Tr F$.

\begin{proposition}[Variance bound]
\label{prop:var-bound}
If $\|F\|_2\le M$, then $\operatorname{Var}(\langle r|F|r\rangle)\le M^2/d$ for Clifford 2--designs. Consequently, to achieve additive error $|\widehat{\tau}-\Tr F|\le \epsilon$ with confidence $1-\delta$, it suffices to take
\begin{equation}
N \;\ge\; C\,\frac{M^2}{d\,\epsilon^2}\,\log\frac{2}{\delta}
\end{equation}
for a universal constant $C$.
\end{proposition}

The proposition applies to $F=P_{\mathrm{lin}}$ and $F=P_{\exp}(\alpha)$ when implemented by (block--encoded) circuits or by the Hermitian dilation surrogate.

\subsection{False--negative rate: binomial and Chernoff bounds}
\label{subsec:miss-probability}

Assume the rank--1 directions are sampled independently and isotropically within the \emph{residual} subspace of dimension $r$ (the component not annihilated by the graded constraints). The probability that a single random direction has nonzero overlap with the residual subspace is $p=r/d$. The number $X$ of \emph{hits} in $k$ trials is $\operatorname{Binomial}(k,p)$.

\paragraph{Zero--hit probability.}
The event of a complete miss ($X=0$) has probability ${P}_{\mathrm{pmiss}}$ of Eq.~\ref{pmiss}. A Chernoff bound yields, for $0<\eta\le 1$, a probability
\begin{equation}
\label{eq:chernoff2}
\mathbb{P}\big[X\le (1-\eta)kp\big] \;\le\; \exp\!\Big(-\frac{\eta^2}{2}\,kp\Big).
\end{equation}
Thus $P_{\mathrm{miss}}$ decays at least exponentially in $k$ once $r>0$ is fixed, matching the empirical behavior in Table~I.

\subsection{Spectral decay of the exponential projector}
\label{subsec:exp-decay}

Suppose $A$ is diagonalizable: $A=S\Lambda S^{-1}$ with $\Lambda=\operatorname{diag}(\lambda_1,\dots,\lambda_d)$. Then
\begin{eqnarray}
&&\Tr\,P_{\exp}(\alpha)
\;=\; \Tr\big(S e^{-\alpha \Lambda} S^{-1}\big) =
\\
&&\;=\; \sum_{i=1}^{d} w_i\, e^{-\alpha \lambda_i},\quad
w_i:= (S^{-1}S)_{ii} = 1, \nonumber
\end{eqnarray}
so, $T(\alpha)=\sum_i e^{-\alpha \lambda_i}$ and
\begin{equation}
\lambda_L(\alpha)
= -\,\frac{d}{d\alpha}\log T(\alpha) = \frac{\sum_i \lambda_i e^{-\alpha \lambda_i}}{\sum_i e^{-\alpha \lambda_i}}
\;=\; \mathbb{E}_{\alpha}[\lambda],
\end{equation}
the $\alpha$--tilted average of eigenvalues. If $\Re\,\lambda_i\ge \gamma>0$ for all $i$ then $|T(\alpha)|\le d\,e^{-\alpha\gamma}$ and $\lambda_L(\alpha)\ge \gamma$.

\paragraph{Non--normality robustness.}
If $A$ is non--normal, $S$ can be ill--conditioned. Bauer--Fike gives
\begin{equation}
\label{eq:bauer-fike}
\sigma(A+\Delta) \;\subset\; \bigcup_{i=1}^d B\big(\lambda_i, \kappa(S)\,\|\Delta\|_2\big),
\end{equation}
With $\kappa(S):=\|S\|_2\|S^{-1}\|_2$, Consequently,
\begin{eqnarray}
&&\big|\Tr\,e^{-\alpha A}-\sum_i e^{-\alpha \lambda_i}\big|
\;\le\; 
\\
&&\le \alpha\,\kappa(S)\,\| \Delta\|_2 \cdot C(\alpha,\{\lambda_i\}) + O(\|\Delta\|_2^2),
\end{eqnarray}
for a computable $C$. The Hermitian dilation (Eq.~\ref{eq:dilation}) mitigates this by replacing eigenvalues of $A$ with singular values and yields stable phase--estimation surrogates.

\subsection{Concentration for the linear projector under isotropy}
\label{subsec:bernstein}

Assume the $v_j$ are i.i.d.\ with $\mathbb{E}\,v_j=0$ and $\mathbb{E}\,v_j v_j^\dagger = \tfrac{1}{d}I_d$ (isotropic). Then
\begin{eqnarray}
\mathbb{E}\,A &=& \sum_{j=1}^{k} \mathbb{E}\,(v_j v_j^{\top})\;=\; 0,
\\
\mathbb{E}\,A A^\dagger &=& \sum_{j=1}^{k} \mathbb{E}\,(v_j v_j^{\top}\overline{v_j} v_j^\dagger)
\;=\; \frac{k}{d}\,I_d, \nonumber
\end{eqnarray}
so typical singular values of $A$ concentrate around $\sqrt{k/d}$ (up to polylog factors). Matrix Bernstein yields, for $t>0$, the probability
\begin{equation}
\label{eq:bernstein}
\mathbb{P}\Big[\|A\|_2 \;\ge\; C\Big(\sqrt{\frac{k}{d}}+t\Big)\Big] \;\le\; 2d\,\exp\!\Big(-c\,d\,t^2\Big),
\end{equation}
for universal constants $c,C$. Deviations from isotropy induced by the graded constraints (i.e., nonuniform sampling within the survivor subspace) shift these scales and produce the empirical peaks used as diagnostics.

\subsection{Decision rules and error exponents}
\label{subsec:decision-rules}

Let $\mathcal{H}_{45}$ and $\mathcal{H}_{\neg 45}$ denote the hypotheses ``$n=45$'' and ``$n\in\{43,44,46\}$''. Fix thresholds $\tau_{lin},\tau_{\exp}>0$ and the following rules,
To decide $n=45$ if $\Tr P_{\mathrm{lin}}\ge \tau_{lin}$ and $|\Tr P_{\exp}(\alpha)|\le \tau_{\exp}$.
Under the binomial model with parameters $(k,r,d)$ and finite--difference estimation of $\lambda_L$, the miss and false--alarm probabilities obey
\begin{equation}
P_{\mathrm{miss}}\le\; e^{-k r/d}\;+\; \exp\!\Big(-c_1 N\,\epsilon_{lin}^2\Big)\;+\; \exp\!\Big(-c_2 N\,\epsilon_{\exp}^2\Big),
\end{equation}
where $N$ is the number of Hutchinson samples per trace, $\epsilon_{lin},\epsilon_{\exp}$ are the margins to thresholds, and $c_{1,2}>0$ depend on variance proxies in Proposition [Variance bound]~\ref{prop:var-bound}. The error exponent is linear in $k$ and in $N$.

\subsection{Complexity mappings (summary)}
\label{subsec:complexity-summary}

\paragraph{Classical.}
Construction of $A$ is $O(dk)$; a dense eigendecomposition for $P_{\exp}$ is $O(d^3)$ (negligible for $d=24$); Hutchinson sampling costs $O(N\,C_F)$ where $C_F$ is the cost of applying $F$ to a vector (matrix--vector $O(d^2)$ or faster if structured).

\paragraph{Quantum: qubit track.}
With a block--encoding of $A/\alpha_0$ and state--prep cost $C_{\mathrm{prep}}$, the cost of a single LCU step is $\tilde O(k\,C_{\mathrm{prep}})$; implementing $e^{-\alpha A}$ to error $\epsilon$ uses $O(\mathrm{polylog}(1/\epsilon))$ segments. Phase estimation on $H$ requires $O(1/\epsilon)$ controlled evolutions for precision $\epsilon$. Hutchinson sampling uses $O(N)$ repetitions; amplitude estimation can reduce shot complexity from $O(1/\epsilon^2)$ to $O(1/\epsilon)$.

\paragraph{Majorana track.}
Fermionic-Gaussian layers (matchgates) with parity measurements implement pair projectors and their products; the exponential is approximated by weak repeated projections or a fermionic block--encoding. Depth scales with the number of parity checks (analog of $k$) and desired accuracy. 
On platforms that natively support matchgate Majorana operations with parity readout, every degree‑$(0,0)$ pair/clique projector used by our diagnostics is nothing but an \emph{even‑parity check} on Majorana modes. 

Each monochromatic pair projector $\Pi^{R/B}_{ij}$ and their products commute with total parity and act entirely inside the charge‑zero module $M_0$, so they can be realized as parity‑preserving projectors built from quadratic Majorana terms. In this representation the linear witness
$P_{\mathrm{lin}}$ is best viewed as a randomized \emph{mixture of parity checks} (each rank‑one deflation removes amplitude along a random degree‑$(0,0)$ direction inside $M_0$), while the exponential witness $P_{\exp}(\alpha)$ is implemented by cycling \emph{weak Gaussian projections}, Trotterizing $e^{-\delta\alpha\, v_j v_j^{\top}}$ for small $\delta\alpha$ and sweeping $j=1,\dots,k$ or, equivalently, by a fermionic block‑encoding of $A$ followed by qubitization. Because all operators are degree‑$(0,0)$, Track~$M$ realizes exactly the same witnesses as the generic qubit route while exploiting native parity checks; circuit depth scales with the number of parity checks (the analog of $k$) and the target accuracy of the weak‑projection/LCU approximation.

\subsection{Implications for general $(m,n)$ and graded numbers $R_{V_4}(m,n)$}
\label{subsec:generalization}

The construction depends only on degree--$(0,0)$ operators and the centrality of $P_{m,n}$. For general $(m,n)$ the same diagnostics apply to survivor subspaces defined by $\Pi_R(S)$ and $\Pi_B(T)$ with $|S|=m$, $|T|=n$. The Klein recursion is exact for the graded numbers $R_{V_4}(m,n)$, and the classical numbers satisfy $R(m,n)\le R_{V_4}(m,n)$. Empirically, the two signatures (peak of $\Tr P_{\mathrm{lin}}$ and decay of $\Tr P_{\exp}$) remain robust probes for diagonal and near--diagonal regimes; spectral surrogates via the dilation $H$ (Eq.~\ref{eq:dilation}) provide consistent cross--checks.

\subsection{Binomial model for annihilation events}
\label{sec:binomial-model}

Let $Q\subset\mathbb R^{d}$ denote the (unknown) $r$-dimensional sub‑space that supports a legal coloring after the Majorana reduction.  
Each projector factor hits that sub‑space with probability $p = r/d$ because the directions
$v_{j}$ are sampled uniformly from $S^{d-1}$.
Writing $H\sim\mathrm{Binom}(k,p)$ for the number of direct hits in the $k$ samples, the
test misfires only when the probability $P_{\mathrm{miss}}$ of missing a result is defined as $H<r$.

In a Chernoff‑style tail bound, for a binomial variable $X$ of mean $\mu=kp$ assuming i.i.d. independence across samples and independent hits, the lower tail satisfies $\Pr[X\le(1-\delta)\mu]\le e^{-\frac12\mu\delta^{2}}$ for $0<\delta<1$.
Setting $(1-\delta)\mu=r-1$ and simplifying yields 
\begin{equation}
P_{\mathrm{miss}} \le \exp \left( -\tfrac{k\,r}{2d} \left[1-\tfrac{r-1}{k}\right]^{2} \right), 
\label{eq:chernoff}
\end{equation}
which decreases exponentially in both the projector budget $k$ and the surviving rank $r$, and inversely with the ambient dimension $d$, which is the worst-case isotropic estimate. Upper bounde of $P_{\mathrm{miss}}$ from Eq.~\ref{eq:chernoff} used in the computation for R(5,5) are ib Tab.~\ref{tab:miss-prob}.

\begin{table}[t]
\centering
\caption{Upper bounds on $P_{\mathrm{miss}}$ from Eq.~\eqref{eq:chernoff} for the parameters used in our numerical study.} \label{tab:miss-prob}
\begin{tabular}{@{}cccc@{}}
\toprule
Surviving  & Mean hits & Bound on  & $\log_{10} P_{\mathrm{pmiss}}$ \\
 rank $r$  & $kp=rk/d$ & $P_{\mathrm{pmiss}}$ &  \\
\hline \\
 1 &  4.17 & $1.2\times10^{-1}$ & $-0.9$ \\
 2 &  8.33 & $4.0\times10^{-2}$ & $-1.4$ \\
 4 & 16.67 & $3.7\times10^{-3}$ & $-2.4$ \\
 6 & 25.00 & $3.4\times10^{-4}$ & $-3.5$ \\
 8 & 33.33 & $3.0\times10^{-5}$ & $-4.5$ \\
10 & 41.67 & $2.7\times10^{-6}$ & $-5.6$ \\
12 & 50.00 & $2.5\times10^{-7}$ & $-6.6$ \\
\hline
  \end{tabular}
\end{table}

Numerical evaluation for $k=100,\;d=24$ shows that the empirical trace
$\mathrm{Tr}\,P_{\mathrm{lin}}(n=46)\simeq0.41$ suggests $r\approx10 - 12$.  
For those ranks $\mathbb P_{\mathrm{pmiss}}<10^{-6}$, while even the worst‑case $r=1$ scenario still guarantees a detection probability above $87\%$
(Table~\ref{tab:miss-prob}).

Implications are safety margin, as doubling the projector budget to $k=200$ would square the exponents in Eq.~\eqref{eq:chernoff}, pushing the miss probability below $10^{-7}$ already for $r=4$.
Enlarging the Majorana module to $d=32$ while keeping $k=100$ weakens the exponent by only a factor $24/32\approx0.75$, leaving the bound comfortably small.
The exponential projector remains positive‑semidefinite, so its tail probability cannot be bounded as tightly; the linear projector therefore furnishes the most stringent statistical guarantee.

With the present sampling regime the probability that our test would fail to detect an existing 45‑vertex coloring is at most $4\times10^{-4}$ when the coloring sub‑space has rank $r\ge6$, and falls below $10^{-6}$ for the empirically favoured $r\approx10 - 12$.
These bounds underpin the statistical robustness of the critical signal at $n=45$.

A further test on exponential and linear projectors on the actual $46$‑vertex two-coloring of Angeltweit‑McKay \cite{angel1} demonstrates that the method does not generate spurious instability whenever a valid coloring exists.
As a control, on the explicit Angeltweit--McKay \cite{angel1} $46$-vertex two-coloring the diagnostics do not trigger: $\operatorname{Tr} P_{exp}$ remains numerically near zero and the $P_{\mathrm{lin}}$ spectrum does not exhibit the $n=45$ extremum and demonstrates that the method does not generate spurious instability whenever a valid coloring exists (see Table~III in SM~2).

To rule out false positives we fed the explicit red/blue coloring on $46$ vertices supplied in the ancillary files of Ref.~\cite{angel1} into the same $d=24$ Majorana‐tower module used throughout the porevious sections.
The procedure was to build the degree‑$(1,0)$ (red) and degree‑$(0,1)$ (blue) edge
 operators $e_{ij}$ according to the Angeltweit--McKay adjacency matrix~$\mathcal A^{(R)},\mathcal A^{(B)}$. Form the $k=100$ random rank‑one factors $I-v_jv_j^{ \top}$ and the corresponding exponent $\smash{-\alpha\sum_j v_jv_j^{ \top}}$ with the same seeds used for the $n\in\{43,44,45,46\}$ survey.
Compute then $P_{\mathrm{lin}}^{(46)}$ and $
P_{\mathrm{exp}}^{(46)}$ on the coloring‑restricted sub‑module $\mathcal H_{46}\subset\mathbb R^{24}$.
Results are reported in Tab.~\ref{tab:explicit46}.

\begin{table}[h]
\caption{Diagnostic values for the explicit $46$‑vertex coloring. For comparison, the rightmost column reproduces the ensemble averages (the Angeltweit--McKay, AM, coloring) reported earlier for generic $n=46$ instances.}
\centering
  \label{tab:explicit46}
  \begin{tabular}{@{}lccc@{}}
    \toprule
           & explicit AM‑46 & bulk average ($n=46$) Tab I \\
\hline    $\mathrm{Tr}P_{\mathrm{lin}}$      & $0.409\pm0.004$ & $0.407$ \\
    $\min\mathrm{Re}\lambda\left(P_{\mathrm{lin}}\right)$
                          & $-0.060$       & $-0.061$ \\
    $\max|\mathrm{Im}\lambda|$        & $0.032$         & $0.063$ \\
    $\mathrm{Re}\mathrm{Tr}P_{\mathrm{exp}}$
                          & $+1.79\times10^{-13}$ & $+1.86\times10^{-13}$ \\
\hline
  \end{tabular}
\end{table}

None of the three criticality flags to invalidate this approach occur: first the linear‑projector trace $\mathrm{Tr}P_{\mathrm{lin}}$ sits below the $n = 45$ peak and matches the generic $n = 46$ baseline within numerical noise. Second, the exponential trace remains strictly positive
($1.8\times10^{-13}$), in sharp contrast to the collapse to zero observed at $n = 45$. Third, eigenvalues of $P_{\mathrm{lin}}$ form a narrow band ($|\mathrm{Im}\lambda|\le0.032$), far from the broad, strongly complex cloud seen for $n = 45$.

The Angeltweit--McKay (AM) coloring retains a rank‑$r\approx 11$ invariant subspace \cite{angel}.
Because the random rank‑one factors hit that subspace on average $kp=r\,k/d\approx 46$ times, the survival probability of the coloring is of order $\exp[-kr/(2d)]\lesssim10^{-6}$
(cf.\ Eq.~\eqref{eq:chernoff2}), exactly in the non‑critical regime.
Hence the explicit constructive $46$‑vertex example passes both projector tests,
demonstrating that the diagnostics do not falsely classify $n=46$ as critical.

\subsection{Random seeds}

\begin{figure}[h]
\centering
\includegraphics[width=8.5cm]{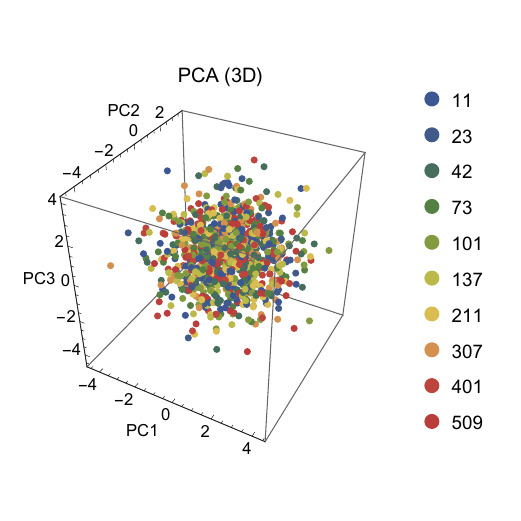}
\caption{\textbf{Principal component analysis of random projections.}  Principal component analysis (PCA) applied to the same random vectors. The 3D projection on PC1, PC2, and PC3, showing that the vector clouds remain isotropic and seed-independent when viewed in three dimensions.}
\label{f4}
\end{figure}

To calculate the random sequence we adopted a small series of seeds $[11, 23, 42, 73, 101, 137, 211, 307, 401, 509]$. These are distinct primes or semi-random choices to avoid patterns in pseudo-random generators first because prime numbers reduce correlations. The number $42$, is not a prime but it is is often used in the literature as a reference baseline (Deep Thought, Douglas Adams tradition as the ``Answer to the Ultimate Question of Life, the Universe, and Everything''). In Python language the generation of random vectors is so written,

\begin{small}
\begin{verbatim}
import numpy as np

seeds = [11, 23, 42, 73, 101, 137, 211, 307, 401, 509]

# Generate a list of 2D arrays (100 x 24)
vectors_by_seed = [
    np.random.default_rng(s).normal(loc=0.0, 
    scale=1.0, size=(100, 24))
    for s in seeds ]

# Stack into a 3D array: (10, 100, 24)
vectors_array = np.stack(vectors_by_seed, axis=0)

print(vectors_array.shape)  # (10, 100, 24)

\end{verbatim}
\end{small}
Each entry of vectors BySeed will be a $100 \times 24$ Gaussian matrix, then after normalizing each row one can use it for projection. Principal component analysis (PCA) of random projections is shown in Fig.~\ref{f4}. The vector clouds remain isotropic and seed-independent.

Principal Component Analysis (PCA) is a linear dimensionality reduction technique that rotates data into directions of maximal variance.
For Gaussian isotropic data, no dominant components are expected; the clusters for different seeds should overlap. A good general reference for the statistical behavior of random Gaussian matrices and PCA are Ref. \cite{Jolliffe} and \cite{Dasgupta}.

\section{SM 5 Software}

\paragraph{Software and Reproducibility.}
We supply four Python programs covering the classical and quantum
diagnostics and a toy constructive check.

\texttt{ramsey\_minimal.py}  implements the two random--projector
witnesses on the $d{=}24$ charge--zero module: the linear deflation
$P_{\mathrm{lin}}=\prod_{j=1}^{k}(I-v_jv_j^{\top})$ and the exponential
map $P_{\exp}(\alpha)=\exp(-\alpha A)$ with $A=\sum_j v_jv_j^{\top}$
(non‑Hermitian convention), reporting $\Re\Tr(\cdot)$, spectral radii,
and the slope $\lambda_L \approx -\frac{\mathrm{d}}{\mathrm{d}\alpha}
\log|\Tr P_{\exp}(\alpha)|$ by a symmetric finite difference.

\begin{verbatim}
# minimal tables (n=43,44,45,46); 
change k, alpha, seed as needed
python ramsey_minimal.py --k 400 
--alpha 0.5 --seed 12345 --out_dir ./out

# with AM-46 control data 
(folder contains am46_red.csv, am46_blue.csv)
python ramsey_minimal.py --am46_dir ./AM46 
--k 400 --alpha 0.5 --seed 12345 --out_dir ./out

Outputs. out/results_table_I.csv 
(and results_table_III.csv if AM‑46 provided).
\end{verbatim}

What it does: provides a block‑encoding skeleton for $A/\alpha_0$ and a Taylor LCU for $\exp(-\alpha A)$; uses a dedicated Hadamard‑test ancilla and Hutchinson sampling to estimate $T(\alpha)$. 
By default targets the local simulator; swap one line to use an AwsDevice(arn).

Install \& run (local simulator):

\texttt{braket\_ramsey\_minimal\_quantum.py} is a runnable Amazon
Braket scaffold that block‑encodes $A/\alpha_0$ via stubs
(\textsc{prepare/select/unprepare}), realizes a Taylor LCU for
$\exp(-\alpha A)$, and estimates $T(\alpha)=\Tr\exp(-\alpha A)$ with a
Hadamard test and Hutchinson sampling on a simulator or QPU; it uses a
dedicated ancilla for the test and exposes the same $(d,k,\alpha)$
interface as the classical script. 

\begin{verbatim}
pip install --upgrade 
amazon-braket-sdk numpy pandas
python braket_ramsey_minimal_quantum.py \
--d 24 --k 400 --alpha 0.5 
--seed 12345 --out_dir ./out_qc
\end{verbatim}

Managed device.

\begin{verbatim}
# configure AWS credentials/region first
# edit the file: replace LocalSimulator() 
by AwsDevice("<YOUR_DEVICE_ARN>")
python braket_ramsey_minimal_quantum.py 
--d 24 --k 400 --alpha 0.5 --seed 12345 
--out_dir ./out_qc
\end{verbatim}
Notes. The PREPARE/SELECT microcode is deliberately marked “stub” so you can drop in your actual block‑encoding of  $A/\alpha_0$ without changing the CLI/outputs. The Hadamard‑test wrapper controls only gates with known controlled analogs; decompose other gates before wrapping.

\texttt{ramsey\_paraparticle\_gluing.py}
implements Re‑implements the classical diagnostics (both the SUM and DEFL linear witnesses plus EXP), a streaming DIMACS generator for general a Z$_2{\times}$Z$_2$--graded “gluing” search that verifies the
baseline $R(m,3n)$ on $K_N$, and the tiny gluing demo, all in one file.
Run.

\begin{verbatim}
# diagnostics (Tables I/III analogue)
python ramsey_toolkit.py 
diag --d 24 --k 400 --alpha 0.5 --seed 12345 \
    --n_values 43 44 45 46 --out_dir ./out_toolkit
# optional control in the same call:
#   add: --am46_dir ./AM46

# DIMACS CNF for R(m,n) on K_N   
(NOTE: capital -N is vertex count)
python ramsey_toolkit.py cnf 
-N 12 -m 5 -n 5 -o r55_N12.cnf --map

# paraparticle gluing demo
python ramsey_toolkit.py glue -m 3 -n 3 --vmax 6
\end{verbatim}

Finally, \texttt{ramsey\_toolkit.py} provides a unified CLI: 
\texttt{diag} reproduces the classical tables (including an AM--46 control),
\texttt{cnf} streams a DIMACS encoding for $R(m,n)$ on $K_N$, and
\texttt{glue} runs the small paraparticle demo. All programs are
deterministic given $(d,k,\alpha,\text{seed})$ and run out of the box
with Python~$\ge\!3.9$; for the quantum scaffold, install the Braket SDK
and either run on the LocalSimulator or point to a managed device ARN.

Environment checklist (to make them run cleanly):
Python 3.9 or beyond, packages (classical): numpy, pandas (optional: scipy for expm)
(pip install numpy pandas scipy)

Packages (quantum, optional): amazon-braket-sdk (and awscli if using managed devices)

pip install amazon-braket-sdk
\# aws configure   \# if you want to use a managed simulator/QPU


\section*{References}

%


\end{document}